\documentclass[%
 reprint,
 amsmath,amssymb, aps,
 prl,
]{revtex4-2}

\usepackage{hyperref}
\hypersetup{
	pdftoolbar=true,        
	pdfmenubar=true,        
	pdffitwindow=false,     
	pdfstartview={FitH},    
	pdftitle={Quantum optimal control robust to $1/f^\alpha$ noises using fractional calculus: voltage-controlled exchange in semiconductor spin qubits},   
	pdfauthor={Bohdan Khromets and Jonathan Baugh}, 
	pdfsubject={Physics -- Quantum Information},  
	pdfnewwindow=true,      
	colorlinks=true,        
	linkcolor=blue,         
	citecolor=green!70!violet,        
	filecolor=magenta,      
	urlcolor=blue           
}
\usepackage{graphicx}
\usepackage{xcolor}
\usepackage{dcolumn}
\usepackage{bm}
\usepackage{physics}
\usepackage{bbm}
\usepackage{amsmath}
 \usepackage{mathrsfs}
\usepackage[mathlines]{lineno}

\begin{document}

\preprint{APS/123-QED}

\title{Quantum optimal control robust to $1/f^\alpha$ noises using fractional calculus: \\ voltage-controlled exchange in semiconductor spin qubits} 

\author{Bohdan Khromets}
 \altaffiliation[Also at ]{Department of Physics, University of Waterloo, 200 University Avenue West, Waterloo, Ontario  N2L 3G1, Canada}
 \email{bohdan.khromets@uwaterloo.ca}
\author{Jonathan Baugh}%
\altaffiliation[Also at ]{Department of Chemistry, University of Waterloo, 200 University Avenue West, Waterloo, Ontario  N2L 3G1, Canada}
 \email{baugh@uwaterloo.ca}
\affiliation{%
Institute for Quantum Computing, University of Waterloo,\\ 200 University Avenue West,  Waterloo, Ontario N2L 3G1, Canada
}%

\date{\today}

\begin{abstract}
	 Low-frequency $1/f^\alpha$ charge noise significantly hinders the performance of voltage-controlled spin qubits in quantum dots. Here, we utilize fractional calculus to design voltage control pulses yielding the highest average fidelities for noisy quantum gate operations. We focus specifically on the exponential voltage control of the exchange interaction generating two-spin $\mathrm{SWAP}^k$ gates. When stationary charge noise is the dominant source of gate infidelity, we derive that the optimal exchange pulse is long and weak, with the broad shape of the symmetric beta distribution function with parameter $1-\alpha/2$. The common practice of making exchange pulses fast and high-amplitude still remains beneficial in the case of strongly nonstationary noise dynamics, modeled as fractional Brownian motion. The proposed methods are applicable to the characterization and optimization of quantum gate operations in various voltage-controlled qubit architectures.

\end{abstract}

\maketitle

\textit{Introduction.---}Quantum dots in semiconductor heterostructures offer precise electric control of highly localized few-electron spin-orbital states. This makes them a promising platform for scalable, spin-based quantum computation. Yet, the electrostatic potential fluctuations in the vicinity of the qubits, known as charge noise, are an important source of gating error and spin dephasing \cite{Culcer2009DephasingSispin, Burkard2023Semiconductorspinqubits} and thus a major threat to the performance and scalability of quantum processors. Charge noise dominates at low frequencies, and its power spectral density is often well approximated over many decades by a power law function $\propto 1/f^\alpha$, with $0< \alpha \lesssim 2$ in experiments
\cite{Chan2018AssessmentEnvNoiseSpectroscopy,Connors2019NoiseSi/SiGe, Kranz2020SingleCrystalMinimizeNoise, Kuhlmann2013ChargeNoiseSpin, Yoneda2017quantumdotspin,
Petit2018ChargeNoiseHotQubits,
Struck2020EnergySplittingNoise_SiSiGe, Jock2022SiliconSinglet–TripletQubit, Connors2022ChargeNoiseSpectroscopy}.
    The background of fluctuating charge traps in the insulator, gate electrode voltage fluctuations (due to interaction with the interface traps, non-ideality of the external electronics, etc.), and white thermal noise, among other things, contribute to the overall incoherent electric noise with the $ 1/f^\alpha$ spectrum \cite{Lundberg2002NoiseSourcesBulkCMOS, Sousa2007Danglingbondspin, Kabytayev2014RobustnessCompositePulsesControlNoise, Shehata2023ModelingChargeEnvironment}.  
    
While ongoing improvements in material growth and device fabrication  \cite{Kranz2020SingleCrystalMinimizeNoise, PaqueletWuetz2023ReducingChargeNoise_QWs, Struck2020EnergySplittingNoise_SiSiGe} allow for the mitigation of decoherence, charge noise cannot be completely eliminated, even in the best-quality devices \cite{Yoneda2017quantumdotspin, Yang2019Siliconqubitfidelities,Mills2022Twoqubitsilicon}. Given the many effects that contribute to noise and the heavily device-dependent noise spectra \cite{RojasArias2023SpatialNoiseCorrelations}, quantum optimal control is necessary to achieve deterministic high-fidelity control with potential for scalability. 
To preserve coherence, dynamical decoupling sequences of strong and fast pulses \cite{Paladino2014ChargeNoiseReview} have been employed in the presence of the Markovian bath \cite{Viola1998DynamicalSuppressionDecoherence, Shiokawa2004DynamicaldecouplingSlowPulses, Uhrig2007KeepingQuantumBitAlive}, and power-law noise in the spin Hamiltonian controls \cite{Faoro2004DynamicalSuppression1/fnoise, Falci2004DynamicalSuppressionTelegraphNoise, Cywinski2008HowEnhanceDephasing, Pasini2010DynamicalDecouplingPowerLawNoise, Ramon2015NonGaussiansignatures}. 
 For single-qubit rotations, time-dependent spin Hamiltonians  decoupled from a single bistable flucutator \cite{Moettoenen2006Highfidelity1QbitGatesChargeNoise, Rebentrost2009OptimalControlNonMarkovianEnv,Pasini2008OptimizationShortCoherent}, or ensemble of such fluctuators giving a power-law spectrum \cite{Kuopanportti2008Suppression1/fnoise}, have been engineered using few-parameter gradient-based optimization. Simultaneous dynamical decoupling and optimal control sequences have been studied for spin Hamiltonians in other systems \cite{Zhang2014ProtectedQuantumComputing,DArrigo2016Highfidelity2qGatesDynamicalDecoupling,Ram2022RobustQuantumControl}. 
Among analytical tools, the geometric formalism \cite{Zeng2018GeneralSolutionInhomogeneousDephasing} has been instrumental in engineering minimum-time quantum gates insensitive to errors up to different orders: single-qubit gates with quasistatic errors \cite{Zeng2018FastestpulsesDynamicallyCorrected_PhaseGates, Zeng2019Geometricformalismconstructing}, pulse and transverse noise errors \cite{Dong2021DoublyGeometricQuantum,Nelson2023DynamicallyCorrectedGates_Multiple_Noise_Sources}, and multiqubit entangling gates \cite{Buterakos2021GeometricalFormalismDynamically}. 

In this Letter, we employ the analytical framework of fractional calculus \cite{samko1993fractional} to design quantum gate operations least sensitive to charge noise when charge noise is the dominant decoherence mechanism (cf. \cite{Hu_2006}). 
 Representing the auto-correlation functions of both stationary and non-stationary $1/f^\alpha$ noise models with fractional integral operators enables us to variationally find smooth pulse shapes yielding highest average gate operation fidelities. 
In addition, we numerically analyze the influence  of pulse shape, length, and noise spectral exponent $\alpha$ on the unitary operation fidelity.  
We focus specifically on the voltage control of exchange interaction $J(V)$---an always positive (and thus non-refocusable) quantity in the absence of the strong out-of-plane magnetic field \cite{Burkard_1999_architecture}---for the generation of $\mathrm{SWAP}^k$ gates on pairs of Loss-Di Vincenzo spin qubits \cite{Loss_1998}. 
Although the idle state of a quantum processor can be chosen as a symmetric (charge noise insensitive) point \cite{Stopa2008MagneticFieldControl, Shim2016ChargeNoiseInsensitive_ExchOnly, Reed2016ReducedSensitivityChargeNoiseSymmetric, Martins2016NoiseSuppressionSymmetricExchangeGates}, entangling operations require sweeping $J(V)$ to noise-sensitive regions. Notably, our framework requires no simplifications of additive and/or quasistatic noise in the spin Hamiltonian. Rather, we fully incorporate the strongly nonlinear dependency of the spin Hamiltonian on voltage controls [such as exponential dependency of $J(V)$] leading to nonlinear noise amplification. This analysis offers more refined strategies for optimal control of exchange in the presence of noise than the standard approach of making pulses as short in time as possible \cite{Nielsen2010Implicationssimultaneousrequirements}, which we show is not necessarily optimal for charge-noise-dominated regimes.

\textit{Exchange gate fidelity.---}Arising from Pauli exclusion principle and Coulomb repulsion, the
isotropic exchange interaction with parameter $J(t)$ for a pair of electron spins $\vec{S}_{1, 2}  = \frac{\hbar}{2} \vec{\sigma}_{1, 2}$ is described by the Heisenberg Hamiltonian: 
\begin{equation}\label{eq:Heisenberg_Ham}
	H(t) = \frac{J(t)}{4} \vec{\sigma_1} \cdot  \vec{\sigma_2} \equiv \frac{J(t)}{2} \left(\mathrm{SWAP}-\frac{1}{2}\right).
\end{equation}
For a system of two spins used as independent qubits, or Loss-Di Vincenzo qubits \cite{Loss_1998}, this Hamiltonian generates a $\mathrm{SWAP}^k$ logic gate operation for the exchange pulse: 
$	J(t) = \pi k \hbar {S(t)}/{T},$
where $T$ is pulse duration, and $S(t)$ is a dimensionless shape function satisfying the normalization condition:
\begin{equation}\label{eq:shape_normalization}
	\left\langle S(t)\right\rangle = \frac{1}{T} \int_{0}^{T} S(t) \dd{t}  = \int_{0}^{1} S(\tau) \dd{\tau} = 1,
\end{equation} 
with $\tau =t/T \in [0,1]$ being the normalized time. 

In realistic quantum processors, such quantum operations as $\mathrm{SWAP}^k$ are driven by increasing the exchange coupling by orders of magnitude from its negligibly small idle value. Such regimes of large control voltage sweeps are characterized by the exponential trend:
$
	J(V) \approx J(V_0) \exp[\varkappa{\left(V-V_0\right)}].
	$
%
Here, $V$ could denote the voltage on the gate electrode controlling the tunneling barrier between the dots \cite{Buonacorsi2021Quantumdotdevices, Shehata2023ModelingChargeEnvironment}, voltage bias between the pair of plunger gate electrodes accumulating electrons underneath \cite{Dial2013ChargeNoiseSpectroscopyExchOscillations, Petta_2005, Connors2022ChargeNoiseSpectroscopy}, or a characteristic of the electric potential landscape such as the tunneling barrier height. 
 The addition of a noisy voltage signal $\widetilde{v}(t)$ with $\left\langle\widetilde{v}\right\rangle=0$ to the ideal pulse $V(t)$
yields the noisy exchange interaction parameter:
$	J(V+\widetilde{v})
	= J(V)  \exp(\varkappa{\widetilde{v}}). $
%
By adjusting $\varkappa$ accordingly, 
 we can combine the contributions of both the voltage fluctuations on gate electrodes and charge trap environment into $\widetilde{v}(t)$. 

As the performance metric of a singular $\mathrm{SWAP}^k$ operation, we choose the average infidelity due to noise \footnote{Note that according to Markov's inequality, 
	average infidelity $\protect\langle\mathscr{F}\protect\rangle$ also gives the upper bound on the extent of the tails of the distribution
	$\protect\mathscr{F}(\protect\widetilde{v})$ 
}. 
For this, we find the overlap of the evolution operators  $U\textbf{[}V(t)\textbf{]}= \exp\left[-\frac{i }{{2\hbar}}\int_{0}^{T} J\textbf{(}V(t)\textbf{)}\dd{t} \mathrm{SWAP} \right]$  in the ideal and noisy cases: 
\(
	\mathcal{F} = \frac{1}{4} 
	\abs{\tr U^\dagger[V] U[V+\widetilde{v}]}\leq 1. 
\)
By expanding for small noises $\widetilde{v}(t)$ and averaging, we find the average infidelity $\left\langle\mathscr{F}\right\rangle = 1-\left\langle \mathcal{F}\right\rangle$:
 \begin{equation}\label{eq:avg_infidelity}
 	\left<	\mathscr{F} \right> \approx \frac{{3}\pi^2}{32} {k^2} \varkappa^2 \int_{0}^T \! \frac{\dd{t_1}}{T} \int_{0}^T \! \frac{\dd{t_2}}{T} S(t_1)S(t_2)	R(t_1,t_2),
 \end{equation}
 where the auto-correlation function of the noise is introduced:
 	$R(t_1,t_2)  = \left<\widetilde{v}(t_1) \widetilde{v}(t_2) \right>.$
 	Expression~\eqref{eq:avg_infidelity} can also be understood as the average gate operation error: the variance of the number of spin swaps $k = \frac{1}{\pi\hbar} \int_{0}^T J(t)\dd{t}$ due to noise, $\left\langle\Delta k^2\right\rangle$, is proportional to $\left<	\mathscr{F} \right>$. 

\textit{Minimization problem.---}
We define the inner product on $[a,b]$:
  $
  \left(\varphi, \psi\right) = \int_a^b \varphi(x) \, \psi(x)\dd{x},
  $
  and associate an operator $\hat{A}$ with a two-variable kernel function $A(x, s)$ when they satisfy $ \qty\big(\hat{A} \varphi ) (x) = \int_a^b A(x,s) \varphi(s) \dd s$. 
In this notation, the optimal pulse design for lowest average infidelity [Eq.~\eqref{eq:avg_infidelity}] is a Lagrangian minimization problem with one Lagrange multiplier $\lambda$ due to the normalization constraint in Eq.~\eqref{eq:shape_normalization}:
  \begin{equation}\label{eq:Lagrangian}
  	\Lambda[S] = (S, \hat{R} S) - \lambda (1,S) \rightarrow \underset{S(t)}{\min{}}.
  \end{equation}
However, it is not obvious that this functional has a minimizer  $S_{\text{opt}}(t)$: this depends strongly on the nature of the noise process.
  We thus proceed to establish the operator representations of auto-correlation $\hat{R}$, proving the existence of $S_{\text{opt}}(t)$ and finding it analytically for relevant models of charge noise. The dependencies of $\left<	\mathscr{F} \right>$ on $T$ and $\alpha$ will be analyzed numerically for certain pulse shapes.

 \textit{Stationary noise.---}
We model the most commonly observed case of stationary charge noise as a  ensemble of independent two-level fluctuators (TLFs), such as distributed interfacial or bulk charge traps \cite{Shehata2023ModelingChargeEnvironment, Sousa2007Danglingbondspin}. The auto-correlation function is assumed to incorporate all relevant mechanisms behind this non-Gaussian noise process. Its general statistical properties (noise ``color'') will be then characterized by the exponent 
$\alpha\!\in\! (0,2)$.
 
A two-level voltage potential fluctuation ${\xi_\gamma (t)}$, with the jumps governed by the Poisson process with the switching rate $\gamma$, has the auto-correlation function:  
	$
 	R_\gamma(t_1,t_2) = \langle\xi_\gamma(t_1)\xi_\gamma(t_2) \rangle = E_\gamma  e^{-2\gamma \left|t_1-t_2\right|},$
 %
and a Lorentzian power spectral density \cite{Yamamoto2017FundamentalsNoiseProcesses,Cywinski2008HowEnhanceDephasing}. $E_\gamma = R_\gamma(t,t)$ will be referred to as the energy of a TLF (measured in $\mathrm{V^2}$).
For an ensemble of TLFs with total energy $R_0$, we consider their energy density per unit range of switching rates of the form \cite{
Milotti2002_1/fNoisePedagogical, Kuopanportti2008Suppression1/fnoise}: 
 \begin{equation}\label{eq:spectral_energy_density}
  {\dd{E(\gamma)}}/{\dd{\gamma}}
 	= {\mathcal{N}^{-1}(\alpha)}
 	\left({R_0}/{\gamma^\alpha}\right)\ \mathbbm{1}_{[\gamma_{\min}, \gamma_{\max}]}.
 \end{equation}
 Here, $\mathbbm{1}_{[a,b]}(\gamma)$ is the indicator function on $[a,b]$ (unity within the interval, and zero outside it), and $\mathcal{N}(\alpha)$ is the normalization parameter.
 In this case, the noise spectral density is well approximated by  $P(f)\approx P_0/\abs{f}^\alpha$ \cite{Milotti2002_1/fNoisePedagogical} in the bulk of the spectrum  $f_{\min} \lesssim  \abs{f}  \lesssim f_{\max}$.
 The spectrum plateaus at $f\lesssim f_{\min}$ and rapidly decays as $ 1/f^2$ at $f \gtrsim f_{\max}$, and the cutoff frequencies are given by
 \(
 f_{\min}={\gamma_{\min}}/{\pi}, \ f_{\max}={\gamma_{\max}}/{\pi}
 \) \cite{Kuopanportti2008Suppression1/fnoise}.
 Note that the relationship between the noise spectral level $P_0=P(\text{1 Hz})$ and its energy $R_0$ depends on the spectral exponent $\alpha$:
 \begin{equation}\label{eq:energy_vs_spectral_level}
 	P_0 = \frac{R_0}{4\sin\frac{\pi\alpha}{2}} \times \begin{cases} 1/\ln(f_{\max}/ f_{\min}), & \alpha=1, \\[5pt]
 		\left({1-\alpha}\right)/{\left( f_{\max}^{1-\alpha} - f_{\min}^{1-\alpha}\right)}, & \alpha\neq 1.
 	\end{cases}
 \end{equation}
Phenomenologically, we relate the low-frequency cutoff to the duration of the experiment $T_{\exp}$, which involves state preparation, quantum operations, and measurements: $ f_{\min} \sim  1/ T_{\exp} $, since all slower frequencies can be calibrated out \cite{Astafiev2004QuantumNoiseJosephson}. The high frequency cutoff is determined by the parasitic capacitances of the device, so $f_{\max}/f_{\min}\gg 1$. 

\begin{figure}[t]
\includegraphics[width=1\linewidth]{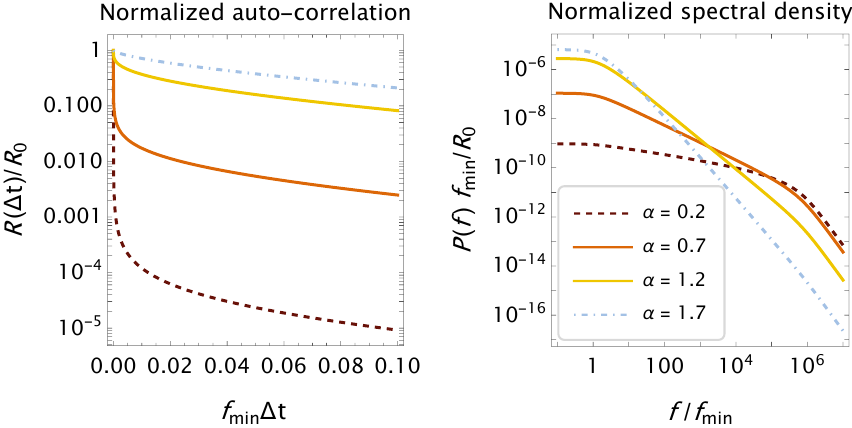}
    \caption{Normalized auto-correlation and power spectral density of the ensemble of TLFs distributed according to Eq.~\eqref{eq:spectral_energy_density}. Here, $\Delta t = \abs{t_1 - t_2}$, and $f_{\max}/f_{\min}=10^6$. 
    }
    \label{fig:autocor_psd}
\end{figure}

The auto-correlation function of the ensemble of TLFs $R(t_1,t_2) = \int  \left({\dd R_\gamma(t_1,t_2)}/{\dd E_\gamma}\right)\left({\dd{E(\gamma)}}/{\dd{\gamma}}\right)\dd\gamma$ can be expressed analytically in terms of the generalized exponential integrals $E_\alpha(z) = \int_1^\infty\dd{s} {e^{-z s}}/{s^\alpha}\ $: 
  \begin{multline}\label{eq:auto-correlation_alpha}
 	\!\!\! R(t_1,t_2) = { 4  P_0} \sin\!\frac{\pi \alpha}{2}  \left[f_{\min}^{1-\alpha} E_{\alpha}(2\pi f_{\min} \abs{t_1-t_2} )  \right. \qquad \\   -\left.{f_{\max}^{1-\alpha}} E_{\alpha}(2\pi f_{\max} \abs{t_1-t_2} ) \right].\!\!\!
 \end{multline}
Figure \ref{fig:autocor_psd} compares the auto-correlation functions [Eq. \eqref{eq:auto-correlation_alpha}] and power spectral densities of stationary $1/f^{\alpha}$ noises with a fixed energy $R_0$, and spectral levels $P_0$ given by Eq.~\eqref{eq:energy_vs_spectral_level}. 
 Exchange pulses are typically short compared to the length of an experiment: $ T/T_{\exp}\ll 1$. Physically, this means that on the timescale $T$ of a single exchange pulse, the noise manifests itself as a random miscalibration of voltage potential with infrequent TLF jumps. Then,  we can use the series expansion of $E_\alpha(z)$ for small $z= 2\pi f_{\min} \abs{t_1-t_2}$ near $z\rightarrow0$ \cite{NistDLMF}:
 \begin{equation}\label{eq:exponential_integral_series}
 	E_\alpha(z) =\! \begin{cases}
 		-C - \ln(z) + O(z),  & \alpha = 1, \\
 	\frac{1}{1-\alpha}\!\left[{z^{\alpha - 1} \Gamma(2- \alpha) - 1}\right]+ O(z), & \alpha \neq 1,
 	\end{cases}
 \end{equation}
where $C\approx 0.577$ is Euler-Mascheroni constant, and $\Gamma(2-\alpha)$ is the gamma function. 
 Note that the the weak singularity in case of $\alpha\in(0,1]$ is integrated out in the expression \eqref{eq:avg_infidelity} for average infidelity. This allows us to drop the last term in the brackets of the auto-correlation formula \eqref{eq:auto-correlation_alpha}, negligible for all significantly nonzero time differences $|t_1-t_2|\gtrsim 1/f_{\max}$,  
 without being concerned about regularization. 
  In Appendix \ref{appendix:autocor_approx}, we improve this approximation  from Eq. \eqref{eq:exponential_integral_series} for longer exchange pulses ($T\sim T_{\exp}$) and show how to minimize the Lagrangian [Eq. \eqref{eq:Lagrangian}] in this case. 
  Additionally, in Appendix \ref{appendix:general_autocor}, we derive the operator representation of auto-correlation and prove the existence of the optimal $\mathrm{SWAP}^k$ exchange pulse shape for an arbitrary ensemble of TLFs.
%

 \textit{Fractional operators.---}   
 To find the operator representation of auto-correlation, we utilize integrals and derivatives of fractional orders \cite{samko1993fractional}. 
Let $L^p(a,b)$ denote a Banach space of Lebesgue-integrable real functions $\varphi: (a,b) \rightarrow \mathbbm{R}$ with finite $p$-norm ($1\leq p < \infty$) \cite{Anastassiou2009FractionalDifferentiationInequalities}.
 Then, for an absolutely integrable $\varphi(x) \in L^1 (a,b)$, the left- and right-sided Riemann-Liouville fractional integral operators of order $\beta>0$, ${}_{a}{I}^\beta $ and ${I}_{b}^\beta$, are defined as follows:
 \begin{equation}\label{eq:RL_integrals_def}
 	\begin{gathered}
 		\left({}_{a}{I}^\beta \varphi\right) (x) = \frac{1}{\Gamma(\beta)} \int_{a}^{x} (x - s)^{\beta - 1} \varphi(s) \dd{s},   \\
 		\left({I}_b^\beta \varphi\right) (x) = \frac{1}{\Gamma(\beta)} \int_{x}^{b} (s - x)^{\beta - 1} \varphi(s) \dd{s}.
 	\end{gathered}
 \end{equation}
 The corresponding kernel functions on $(a,b)$ can be written in terms of indicator functions, for instance:
 \begin{equation}\label{eq:RL_kernel_functions}
	{}_{a}I^\beta(x,s) = \left[{(x-s)^{\beta-1}}/\,{\Gamma(\beta)}\right]\mathbbm{1}_{[a,x]}(s),
 \end{equation}
 and similarly for $I_b^\beta(x,s)$. 
 Integrals \eqref{eq:RL_integrals_def} have inverse operators known as the Riemann-Liouville fractional derivatives. For functions $\varphi(x)$ with ${}_{a}I^{\beta}\varphi \in L^1(a, b)$ or $I_b^{\beta}\varphi \in L^1(a, b)$, respectively, they are defined as combinations of ordinary differentiation and fractional integration:
 \begin{equation}\label{eq:RL_derivs}
 \begin{gathered}
 	{}_{a}D^{\beta} \varphi=\left({\dd{}}/{\dd{x}}\right)^{\![\beta] + 1} {\!\!}_{a}I^{1 - \{\beta\}}\varphi,
 	\\ 
 	D_b^{\beta} \varphi=  \left(- {\dd{}}/{\dd{x}}\right)^{\![\beta] + 1}\!\! I_b^{1 - \{\beta\}} \varphi,
 \end{gathered} 
 \end{equation}
 where $[\beta]\in\mathbb{Z}$ and $\{\beta\}\in[0,1)$ are the integer and fractional parts of $\beta$, respectively: 
 $\beta = [\beta] + \{\beta\}$.
 Importantly, ${}_{a}I^{\beta} $ and $I_{b}^{\beta} $ are conjugate operators \cite{abdeljawad2019by_parts} with respect to the inner product:
\begin{equation}\label{eq:int_conjugate}
 	\qty\big( \varphi ,\ {}_{a}I^\beta\psi ) = 	\qty\big( I_b^\beta\varphi ,\ \psi ), \qquad \left({}_{a}I^{\beta}\right)^* = I_{b}^{\beta},
 \end{equation}
 for the properly Lebesgue integrable $\varphi, \psi$ \footnote{
  The conjugation relation \protect\eqref{eq:int_conjugate} for $\protect\beta>0$ holds for $\protect\varphi(x) \in L\protect^p(a, b)$, $\protect\psi(x) \protect\in L\protect^q(a, b)$ satisfying $\ p, q \protect\geq 1$, and ${p}^{-1} + {q}^{-1} \protect\leq 1 + \protect\beta $; in case of equality, $p, q \protect\neq 1$.}.
    For functions $\varphi(x)$ with certain regularity conditions near the boundaries such as $ I_b^{1-\beta} \varphi \overset{x\rightarrow b}{\longrightarrow}0$, the conjugation rule for derivatives holds as well: $\left({}_{a}D^\beta\right)^* = D_b^\beta$ \cite{abdeljawad2019by_parts}.

Riemann-Liouville operators straightforwardly generalize the conventional derivatives and integrals of power functions: left-sided operations that act on $(x-a)^\kappa$ and right-sided ones that act on $(b-x)^\kappa$. For example, ${\ }_{a}D^\beta (x -a )^\kappa = \frac{\Gamma(\kappa + 1)}{\Gamma(\kappa - \beta + 1)} (x -a )^{\kappa - \beta}.$
 
\textit{Optimal shapes.---}
\begin{figure*}
\includegraphics[width=0.95\linewidth]{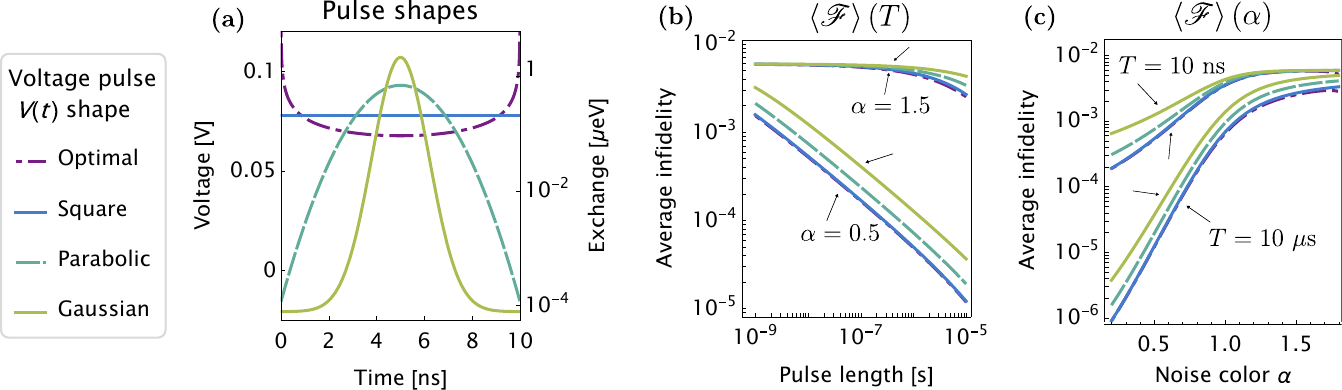}
	\caption{Quantum operation performance in a stationary $1/f^\alpha$ noise environment. (a) Voltage/exchange pulses of four distinct shapes that generate a $\mathrm{SWAP}$ operation, shown for $T=10$ ns. The optimal pulse shape from Eq.~\eqref{eq:opt_voltage_pulse} is shown for $\alpha=1.4$.
		(b) Average infidelity as a function of pulse length for two values of $\alpha$.
		(c)  Average infidelity as a function of $\alpha$ for two values of pulse length. In (b) and (c), frequency cutoffs and noise energy are set to $f_{\min}=10$ kHz, $f_{\max}=10$ GHz, $R_0 =1$ mV$^2$.}
	\label{fig:infid_dependency}
\end{figure*}
 For $\alpha\neq 1$, we now use formulae \eqref{eq:auto-correlation_alpha} and \eqref{eq:exponential_integral_series} to write the dominant contribution to auto-correlation in terms of normalized times $\tau_{1,2}=t_{1,2}/T\in[0,1]$, and a small parameter $\theta = 2 \pi f_{\min}T \sim T/T_{\exp} \ll 1 $:
 \begin{equation}\label{eq:auto-correlation_a<1}
 	R(\tau_1, \tau_2) \approx 
 	\frac{4 P_0\sin\!\frac{\pi\alpha}{2}f_{\min}^{1-\alpha}}{1-\alpha}
 	\qty\Bigg[ \frac{\theta^{\alpha - 1} \Gamma({2-\alpha}) }{\abs{\tau_1 -\tau_2}^{1 -\alpha}}  - 1]. 
 \end{equation}
For $\alpha\in(0,1)$, the expression ${\abs{\tau_1 -\tau_2}^{\alpha-1} }$ has a symmetric integral representation \cite{williams1963class_int_eq}:
 \begin{equation}\label{eq:tau_diff_operator_form<1}
 {\abs{\tau_1 -\tau_2}^{\alpha-1}} = 
  \frac{\pi \int_{0}^{1} \dd{s} K(\tau_1, s)K(\tau_2, s) \dd{s}}{\Gamma\left( 1 -{\alpha}\right)\sin\frac{\pi\alpha}{2}},
 \end{equation}
 with 
$ 	K(\tau, s) = \frac{\tau^{\alpha/2}}{\Gamma\left({\alpha}/{2}\right)}(\tau - s)^{{\alpha}/{2}-1}\; s^{-\alpha/2}\ \mathbbm{1}_{[0,\tau]}(s).$
 %
Formula \eqref{eq:RL_kernel_functions} suggests that $K\left({\tau, s}\right)$ has a fractional operator representation:
 $	\hat{K} =\tau^{\alpha/2}\; {}_{0}I^{\alpha/2} \tau^{-\alpha/2} 
 . 	
 $
 %
Thus, ${\abs{\tau_1 -\tau_2}^{\alpha-1}}$ corresponds to a self-adjoint operator proportional to $ \hat{K}\hat{K}^{*}.$ 
 Introducing $M = \hat{K}^{*}S =\tau^{-\alpha/2}  I_1^{\alpha/2} \tau^{\alpha/2} S$, 
 we obtain the Lagrangian from Eq. \eqref{eq:Lagrangian}:
 \begin{equation*}\label{eq:Lagrangian_a<1}
 	\Lambda[M] \propto  
 	\qty\bigg[ (M, M)  - \frac{\theta^{1-\alpha}\sin\frac{\pi\alpha}{2}}{\pi(1-\alpha)} ]-\lambda (\hat{K}^{-1} 1, M),
 \end{equation*} 
 which is a quadratic, positive-definite functional of $M$.
From ${\delta\Lambda}/{\delta M }= 0$, we thus obtain the minimizer $M_\text{opt}(\tau) \propto \tau^{\alpha/2} {\,}_{0}D^{\alpha/2}\left(\tau^{-\alpha/2} \right) \propto \tau^{-\alpha/2}$. Finally, using $S =\tau^{-\alpha/2} D_1^{\alpha/2}\tau^{\alpha/2} M$ and normalizing to $(1, S)=1$, we determine the
optimal shape of exchange pulse: 
\begin{equation}\label{eq:opt_shape}
	S_\text{opt}(\tau) = \frac{\tau^{-{\alpha}/{2}} (1-\tau)^{-{\alpha}/{2}} }{B\left( 1-\frac{\alpha}{2}, 1-\frac{\alpha}{2}\right) } ,
\end{equation}
where $B(a,b)= \Gamma(a)\Gamma(b)/\Gamma(a+b)$ is beta function. 

In case of $\alpha> 1$, we utilize the expression originally derived by Decreusefond and \"Ust\"unel \cite{Decreusefond1999StochasticAnalysisFBM} for fractional Brownian motion, a non-stationary stochastic process that will be briefly discussed later:
\begin{equation}\label{eq:tau_diff_operator_form_alpha>1}
	\begin{aligned}
		- \abs{\tau_1- \tau_2}^{\alpha -1} &= \frac{\pi(\alpha-1)\hat{K}_\textsc{fbm}^{\,}\hat{K}_\textsc{fbm}^{*}}{\Gamma(2-\alpha)\sin\frac{\pi\alpha}{2}}  - 	   \left(\tau_1^{\alpha -1} + \tau_2^{\alpha -1}\right), \\
		\hat{K}_\textsc{fbm}^{\,} &= {}_0 I^{\alpha - 1} {\,}\tau^{1-{\alpha}/{2}} {{\,}}_0 I^{1- {\alpha}/{2}} {\,}\tau^{{\alpha}/{2} -1}.
	\end{aligned}
\end{equation}
Just as above, the Lagrangian from Eq. \eqref{eq:Lagrangian} becomes a positive-definite functional of $M = \hat{K}_\textsc{fbm}^{*}S$, and the same minimization procedure applies.
We derive in Appendix \ref{appendix:opt_shape} that the optimal shape $S_\text{opt}(\tau)$ is given by the same expression \eqref{eq:opt_shape} for $1/f^\alpha$ noises with any $\alpha\in(0,2)$. 
To show this in the limiting case of $\alpha=1$, we utilize the Chebyshev polynomial basis expansion of auto-correlation instead of fractional calculus.

\textit{Numerical analysis.---} 
We obtained analytically that the shape of a voltage-controlled exchange pulse $J\textbf{(}V(t)\textbf{)}$  giving optimal performance in the presence of stationary $1/f^\alpha$ noise is a symmetric beta distribution function with parameter $1-\alpha/2$ [Eq.~\eqref{eq:opt_shape}]. The corresponding voltage pulse $V(t)=V_0 + \ln\left[J(t)/J(V_0)\right]/\varkappa$ then becomes:
\begin{eqnarray}\label{eq:opt_voltage_pulse}
		V_\text{opt}(t) 
		= V_0 - \tfrac{1}{\varkappa} \ln\left[{J(V_0) T\, B\!\left( 1-\tfrac{\alpha}{2}, 1-\tfrac{\alpha}{2}\right)}/{\pi k \hbar }\right]\qquad&\nonumber\\  -
   \left({\alpha}/{2\varkappa}\right) \,\ln\!\left[\tfrac{t}{T}\left(1 - \tfrac{t}{T}\right) \right]\!.\ \qquad&
\end{eqnarray}
Experimentally, the voltage pulse would be a non-singular approximation to this function, and ensure a small idle value of exchange before and after each pulse $J(t=0, T)\ll \hbar/ T$.

The numerically calculated dependencies of average infidelity [Eq. \eqref{eq:avg_infidelity}] on pulse length, shape, and noise spectral exponent $\alpha$ are given in Figure~\ref{fig:infid_dependency}. We utilize the exact expression for auto-correlation \eqref{eq:auto-correlation_alpha}, assuming the same total energy $R_0$ for all noise processes with different $\alpha$. 
The amplitudes and widths of all pulses shown in Fig.~\ref{fig:infid_dependency}(a) are adjusted to perform a $\mathrm{SWAP}$ gate operation with $k=1$.
For example, a Gaussian shape of exchange pulse $S(\tau) \propto \exp({-(\tau-1/2)^2/2\sigma^2})$ corresponds to a parabolic voltage pulse, and the exponential-of-Gaussian exchange $S(\tau)\propto \exp\left[ h \exp({-(\tau-1/2)^2/2\sigma^2})\right] $ is realized with a Gaussian voltage pulse. We fix $\sigma=0.12, h=10$ henceforth. 
Device-specific parameters are adapted from the full-configuration-interaction calculation for a Si/SiO$_2$ double quantum dot with a tunneling gate voltage control of exchange \cite{Buonacorsi2020Optimizinglateralquantum}: $V_0=0.04 $ V, $J(V_0)=0.01\ \mu$eV, $\varkappa=80 $ V$^{-1}$. Gates with other values of $k$ such as $\sqrt{\mathrm{SWAP}}$ with $k=\frac{1}{2}$ are straightforward to analyze in the same fashion due to the relation $\left<\mathscr{F}\right>\propto k^2$, independent of the noise model [Eq. \eqref{eq:avg_infidelity}]. 

It is found that the optimal voltage pulses from Eq. \eqref{eq:opt_voltage_pulse} perform almost identically to square pulses. However, either shape reduces the average infidelity by up to a factor of 4 when compared to a Gaussian voltage pulse. 
The most important trend evident in Fig.~\ref{fig:infid_dependency}(b) is the decrease in $\left<\mathscr{F}\right>$ with the duration of the pulse, reaching a rate of $-5.2 $ dB/decade for $\alpha=0.5$.
 It is consistent with the approximate expressions for auto-correlation: $A\ \mathrm{sign}(1-\alpha)(B\, T^{\alpha -1} -1) $ for $\alpha\neq 1$ [Eq. \eqref{eq:auto-correlation_a<1}], or $A - B\, \ln(T)$ for $\alpha=1$ [Eq. \eqref{eq:quadratic_form_Chebyshev}] for some $A, B>0$, all of which are monotonically decreasing functions of $T$. 
 This trend goes against the common strategy of making exchange pulses as short in time and high in amplitude as possible to minimize the impact of noise
\cite{Nielsen2010Implicationssimultaneousrequirements}. 
Our results therefore suggest an alternative strategy of using long, low-amplitude, and broad pulses: the exponential suppression of noise when $J(t)$ is kept low throughout the pulse may outweigh the loss in fidelity due to background decoherence processes. Experimentally, we anticipate a ``sweet spot'' in pulse length. As $T_2$ values increase due to material and fabrication improvements, this strategy is expected to become more relevant. 

As follows from Figs. \ref{fig:infid_dependency}(b, c), the gain in fidelity from the optimal choice of pulse length and shape is most significant for weakly correlated noise environments, i.e., with small values of $\alpha$. Such spectral behavior is observed, for example, when the cotributions of white and $1/f$ flicker noises are both significant. 

\textit{Comparison with non-stationary model.---}
 We now investigate how a $\mathrm{SWAP}^k$ gate performance is influenced by the non-stationarity of the noise process, which we  choose to model as the fractional Brownian motion (fBm)
 \cite{mandelbrot1968fractional}.
 This process $b_\alpha(t)$ is characterized by $b_\alpha(0)=0$, statistical independence of increments, and auto-correlation function: 
  \begin{eqnarray}\label{eq:auto-correlation_FBM}
 	R_\textsc{fbm}(\tau_1, \tau_2) =2P_0&(2\pi T)^{\alpha-1}   \abs{\Gamma(1-\alpha) \sin\frac{\pi\alpha}{2} }\qquad\nonumber\\ \times &\left( \tau_1^{\alpha -1} + \tau_2^{\alpha -1} - \abs{\tau_1- \tau_2}^{\alpha -1} \right)\!.\quad
 \end{eqnarray}
 In the generalized sense of a  Wigner-Ville spectrum $P(t,f)$ \cite{flandrin1989spectrum, kuleshov2013spectral}, fBm  demonstrates a desired spectral behavior $P_0/|f|^\alpha$ asymptotically for $t\rightarrow\infty$.
 Physically, the fBm model could be applicable to devices with a high density of charge traps, where charge transport was previously modeled with fractional differential equations \cite{Sibatov2007FractionalDifferentialKinetics,Alaria2019ApplicationFractionalOperators}.

 As shown in \cite{Decreusefond1999StochasticAnalysisFBM}, expression \eqref{eq:auto-correlation_FBM}
 has an operator representation $\hat{R}_\textsc{fbm} = (2\pi)^{\alpha} P_0\, T^{\alpha-1}  \hat{K}_{\textsc{fbm}} \hat{K}_{\textsc{fbm}}^{*}
$, where $\hat{K}_{\textsc{fbm}}$ for the subdiffusive fBm with $\alpha\in(1,2)$ is defined in Eq.~\eqref{eq:tau_diff_operator_form_alpha>1}.
The Lagrangian [Eq. \eqref{eq:Lagrangian}] then becomes a quadratic functional of $M = \hat{K}_{\textsc{fbm}}^{*} S$: 
$\Lambda[M]\propto (M, M) - \lambda (\hat{K}_{\textsc{fbm}}^{-1} 1, M),$
with a variational minimizer given by the generalized function $M_\text{opt}(\tau) = \lambda' \tau^{1-\alpha/2}\delta(\tau-0^+)$.
This indicates that the normalized shape of exchange pulse must be delta-like: $ S_\text{opt} =\delta(\tau-0^+)$.

This result can be understood heuristically. Since $R_\textsc{fbm}(\tau_1, \tau_2)$ is growing monotonically with respect to both variables, the effect of noise on fidelity will progressively worsen over time. Thus, making the pulse as localized near $t=0$ as possible ensures minimal impact: $( S_\text{opt}, \hat{R}_\textsc{fbm} S_\text{opt})\rightarrow 0,$ [cf. Eq.~\eqref{eq:auto-correlation_FBM}]. Likewise, the pulse length should be made as short as possible for optimal performance, which is consistent with $(S, \hat{R}S )\propto T^{\alpha-1}$. 

\textit{Conclusions.---}
 In summary, we address the problem of voltage pulse design for quantum dot spin qubits in the presence of $1/f^\alpha$ charge noise numerically and analytically, using the framework of fractional calculus. 
 Long, low-amplitude, and broad pulses on average yield significantly higher fidelities of $\mathrm{SWAP}^k$ exchange gates in the presence of stationary noises, especially those with small $\alpha$ values indicating weakly correlated noise environments. This is consistent with the result from \cite{Shiokawa2004DynamicaldecouplingSlowPulses} that dynamical decoupling need not contain ultrafast pulses for optimal performance. The common choice of short, high-amplitude pulses is thus advantageous only in case of a strongly non-stationary noise environment or systems with short coherence times.  
 Strong exchange coupling regimes ($\gtrsim 10^{-5}$ eV) thus may not be necessary for high-fidelity quantum operations. Finally, although we focused on the exchange-controlled two-spin $\mathrm{SWAP}^k$ gates, the fractional operator representation method presented in this work is equally applicable to engineering other spin qubit Hamiltonians controlled with noisy voltage signals. These could include singlet-triplet or exchange-only spin qubits, where $J\textbf{(}V(t)\textbf{)}$ pulses realize other quantum operations on logical qubits such as single-qubit rotations, or other effective spin couplings, arising, for example, from spin-orbit interaction.

\begin{acknowledgments} 
This research was undertaken thanks in part to funding from the Canada First Research Excellence Fund (Transformative Quantum Technologies) and the Natural Sciences and Engineering Research Council (NSERC) of Canada.
\end{acknowledgments}

\appendix
\renewcommand{\thesection}{\Alph{section}}
\section{Appendix A. Auto-correlation operator of a general stationary charge noise }\label{appendix:general_autocor}
Consider an ensemble of two-level fluctuators (TLFs) with an arbitrary energy density  $\frac{\dd{E(\gamma)}}{\dd{\gamma}}\geq0$ due to a distribution of switching rates. 
We aim to express the auto-correlation function of such a noise process:
  \begin{equation}\label{eq:auto_correlation_arb_dist}
  	R(t_1,t_2) = \int_{0}^{\infty} \dd{\gamma} \frac{\dd{E(\gamma)}}{\dd{\gamma}}\; e^{-2\gamma \abs{t_1 - t_2}}, \tag{A1}
 \end{equation}
 in terms of the non-singular Caputo-Fabrizio fractional integral operators \cite{caputo2016applications}: 
\begin{equation}\label{eq:caputo-fabrizio_integrals}
	\begin{gathered}
		\left({}_{a}\mathscr{I}^{\beta} \varphi\right)(t) 
		= \frac{1}{\beta} \int_{a}^{t}  \varphi(s) \exp\left[-\tfrac{1-\beta}{\beta}(t-s)\right] \dd{s}, \\
		\left(\mathscr{I}_{b}^{\beta} \varphi\right)(t) 
	= \frac{1}{\beta} \int_{t}^{b}  \varphi(s) \exp\left[-\tfrac{1-\beta}{\beta}(s-t)\right] \dd{s}.
	\end{gathered}
\tag{A2}
\end{equation}
The fractional derivative operators \cite{losada2015inverse}:
\begin{equation}\label{eq:caputo-fabrizio_derivatives}
{\,}_{a}\mathscr{D}^{\beta} 
	= \beta \frac{\dd{}}{\dd{t}}+ (1 - \beta), \qquad \mathscr{D}_{b}^{\beta} = - \beta \frac{\dd{}}{\dd{t}} + \left(1 - \beta\right), \tag{A3}
\end{equation}
are connected to the corresponding fractional integrals as follows:
\begin{equation*}\label{eq:caputo_fabrizio_inverse}
	\begin{gathered}
		\left({}_{a}\mathscr{D}^{\beta} {\,}_{a}\mathscr{I}^{\beta} \varphi\right)\!(t) = \varphi(t), 
		\\ 
		\left({}_{a}\mathscr{I}^{\beta} {\,}_{a}\mathscr{D}^{\beta} \varphi\right)\!(t)
		= \varphi(t) - \varphi(a) \exp\left[-\tfrac{1-\beta}{\beta}(t-a)\right].
	\end{gathered}
\end{equation*}
As in the Riemann-Liouville case, left and right-sided Caputo-Fabrizio integrals are conjugate operators: $ \left({}_{a}\mathscr{I}^{\beta} \right)^* = \mathscr{I}_{b}^{\beta}$.

We use the identity $\min(t_1, t_2) \equiv \frac{1}{2}\left(t_1 + t_2 - \abs{t_1 - t_2}\right)$ to express the exponential difference kernel as follows:
\begin{equation}\label{eq:exponential_repr}
	e^{-2\gamma \abs{t_1 - t_2}} = e^{-2\gamma (t_1 + t_2)} + 4 \gamma\!\!\!\!\!\!\!\!\!\! \int\displaylimits_{0}^{\quad\min(t_1, t_2)}\!\!\!\!\!\!\!\!\! \dd{s} e^{-2\gamma(t_1 - s) }  e^{-2\gamma(t_2 - s)} .\tag{A4}
\end{equation}
This expression contains an inner product on $[0,T]$ of the functions of type $ e^{-2\gamma(t - s)}\; \mathbbm{1}_{[0, t]}(s)$ . As follows from Eq. \eqref{eq:caputo-fabrizio_integrals}, these terms are proportional to the kernel functions of Caputo-Fabrizio integrals with $\frac{1-\beta}{\beta}=2\gamma$, or  ${\beta}=1/(2\gamma+1)$. 
The generalized auto-correlation from Eq. \eqref{eq:auto_correlation_arb_dist} thus corresponds to a positive definite, self-adjoint operator on $[0,T]$:
\begin{multline}\label{eq:auto-correlation_op_repr_CF}
		\hat{R}= \int_{0}^{\infty} \dd{\gamma} \left\{\frac{\dd{E(\gamma)}}{\dd{\gamma}}
		\right.
		\\\times \left. \left[
		 e^{-2\gamma (t_1 + t_2)} + \frac{4\gamma}{(2\gamma + 1)^2}{\, }_{0}\mathscr{I}^{\frac{1}{2\gamma +1}}\left( {\, }_{0}\mathscr{I}^{\frac{1}{2\gamma +1}}\right)^{*} \right]\right\}. \tag{A5}
\end{multline}
 The optimal $\mathrm{SWAP}^k$ exchange pulse shape $S(t)$ can then be found from the Lagrangian [Eq. \eqref{eq:Lagrangian}]:
\begin{multline}\label{eq:Lagrangian_general_dist}
	\Lambda[S]  =  -\lambda (1, S) + \int_{0}^{\infty} \dd{\gamma}\left\{ \frac{\dd{E(\gamma)}}{\dd{\gamma}}\right. \\\times\left.
	\left[ \left(S, e^{-2\gamma t}\right)^2 + \frac{4\gamma}{(2\gamma + 1)^2}\left(\mathscr{I}_T^{\frac{1}{2\gamma +1}}S, \mathscr{I}_T^{\frac{1}{2\gamma +1}}S\right) \right]\right\}, \tag{A6}
\end{multline}
which clearly has a nonnegative second variation and thus a minimum.
This proves the existence of the optimal exchange pulse shape realizing a $\mathrm{SWAP}^k$ gate in any stationary noise environments due to ensembles of TLFs.

\section{Appendix B. Optimal exchange pulse shape for $1/f^\alpha$ noises with $\alpha\geq 1$}\label{appendix:opt_shape}

 \textit{Case of ${\alpha>1}$.---}
 In the main text, we established the self-adjoint operator form [Eq. \eqref{eq:tau_diff_operator_form_alpha>1}] of auto-correlation function $R(\tau_1, \tau_2)$ from Eq. \eqref{eq:auto-correlation_a<1}. Together with the normalization condition $\left(1, S\right)=1$, it yields a positive-definite Lagrangian:
\begin{multline}\label{eq:bilinear_form_alpha>1}
\Lambda[M]	\propto - \lambda(1, S) +
	\frac{\sin\frac{\pi\alpha}{2} }{\pi(\alpha -1)} + \left\{ \left( 2\pi f_{\min} T\right)^{\alpha-1}\right. \\ \times\left.
\left[\left(M, M\right) - \frac{2\Gamma({2-\alpha}) \sin\frac{\pi\alpha}{2} }{\pi(\alpha-1)}\left(\tau^{\alpha -1}, S\right) \right]\right\} , \tag{B1}
\end{multline}
where $M$ and $S$ are connected via the kernel operator of fractional Brownian motion (fBm):
\begin{equation}\label{eq:S_and_M_alpha>1}
	S =\left( \hat{K}_\textsc{fbm}^{*}\right)^{\!-1}\!\!M\ = D_1^{ \alpha-1} \tau^{\frac{\alpha}{2} -1} D_1^{1-\frac{\alpha}{2}}\tau^{1-\frac{\alpha}{2}} M. \tag{B2}
\end{equation}
The extremum condition ${\delta \Lambda}/{\delta M}=0$ yields:
\begin{eqnarray*}\label{eq:extremum_alpha>1}
	M_\text{opt}(\tau) = &\frac{\Gamma({2-\alpha}) \sin\frac{\pi\alpha}{2} }{\pi(\alpha-1)}\tau^{1-\frac{\alpha}{2} }  D_{0}^{1 - \frac{\alpha}{2}} \tau^{\frac{\alpha}{2}-1} D_{0}^{\alpha-1} \tau^{\alpha -1} \nonumber\\
&	+   {\lambda'}\, \hat{K}_\textsc{fbm}^{-1}  1 = \frac{\Gamma\left(2-\alpha\right)}{\Gamma\left(1-\frac{\alpha}{2} \right)} \tau^{\frac{\alpha}{2} -1}+   {\lambda'}\, \hat{K}_\textsc{fbm}^{-1}  1 ,
\end{eqnarray*}
for some rescaled Lagrange multiplier $\lambda'$. The first term was simplified using the identities ${\ }_{0}D^\beta x^\kappa = \frac{\Gamma(\kappa + 1)}{\Gamma(\kappa - \beta + 1)} x^{\kappa - \beta},$ and $\Gamma(p)\Gamma(1-p)=\frac{\pi}{\sin\pi p}$. Using Eq. \eqref{eq:S_and_M_alpha>1}, we obtain the optimal shape of exchange pulse:
\begin{equation*}\label{eq:S_alpha>1_derivation}
	S_\text{opt}(\tau) =  \tfrac{\Gamma\left(2-\alpha\right)}{\Gamma\left(1-\frac{\alpha}{2} \right)} D_1^{ \alpha-1} \tau^{\frac{\alpha}{2} -1} D_1^{1-\frac{\alpha}{2}}1 +  {\lambda'}\left(\hat{K}_\textsc{fbm}\hat{K}_\textsc{fbm}^{*}\right)^{\!-1\!}\!\!  1.
\end{equation*}
We now use $D_1^{1-\frac{\alpha}{2}}1 = \frac{1}{\Gamma({\alpha}/{2})}(1-\tau)^{\frac{\alpha}{2}-1}$ and
 write the first term of the expression above explicitly as a differintegral: 
\begin{gather*}\label{eq:S_alpha>1_integral}
	\tfrac{\Gamma\left(2-\alpha\right)}{\Gamma\left(1-\frac{\alpha}{2} \right) \Gamma\left(\frac{\alpha}{2} \right)} D_1^{ \alpha-1}\tau^{\frac{\alpha}{2}-1} (1-\tau)^{\frac{\alpha}{2}-1} = \\
	-\tfrac{1}{\Gamma\left(1-\frac{\alpha}{2} \right) \Gamma\left(\frac{\alpha}{2} \right)} \frac{\dd}{\dd\tau} \int_{\tau}^{1} \dd{s} (s-\tau)^{1-\alpha} s^{\frac{\alpha}{2} -1}(1-s)^{\frac{\alpha}{2} -1}.
\end{gather*}
With the change of variables $\xi = \frac{s-\tau}{1-\tau}$, the expression above reduces to the standard integral representation of the hypergeometric function $F(a,b;c; z)$ \cite{NistDLMF}: 
\begin{equation*}\label{eq:S_alpha>1_hypergeometric}
		-\frac{\Gamma\left(2-\alpha\right) \frac{\dd}{\dd\tau} \left[ \left(\tfrac{1-\tau}{\tau}\right)^{1-\frac{\alpha}{2}}  F\left(1-\tfrac{\alpha}{2}, 2-\alpha; 2-\tfrac{\alpha}{2}; \tfrac{\tau-1}{\tau}\right) \right]}{\Gamma\left(1-\frac{\alpha}{2}\right)\Gamma\left(2-\frac{\alpha}{2}\right)}.
\end{equation*}
Since   $c=a+1$ is satisfied for the hypergeometric function above, it is related to the incomplete beta function $B(x; p, q) $  \cite{NistDLMF}: 
\begin{equation*}\label{eq:beta_hypergeometric}
B(x; p, q) 	= \frac{x^p}{p} F(p, 1-q; p+1; x).
\end{equation*}
The first term of $S(\tau)$ thus reads as follows:
\begin{equation*}\label{eq:S_alpha>1_beta}
		\frac{\Gamma\left(2-\alpha\right) \left(1-\frac{\alpha}{2}\right)e^{-{i\pi\alpha}/{2}}}{\Gamma\left(1-\frac{\alpha}{2}\right)\Gamma\left(2-\frac{\alpha}{2}\right)}  \frac{\dd}{\dd{\tau}} B \left(\frac{\tau-1}{\tau}; 1-\frac{\alpha}{2}, \alpha-1\right)\!.
\end{equation*}
The integral definition of $B(x; p, q)\equiv\int_0^x t^{p-1} (1-t)^{q-1}\dd{t}$ is straightforward to differentiate, and the pre-factor can be simplified using the basic properties of gamma functions. This yields the optimal shape function:
\begin{equation}\label{eq:S_alpha>1_final}
S_\text{opt}(\tau) = \frac{\tau^{-\frac{\alpha}{2}} (1-\tau)^{-\frac{\alpha}{2}}}{B\left(1-\frac{\alpha}{2}, 1-\frac{\alpha}{2}\right)}   +  {\lambda'}\left(\hat{K}_\textsc{fbm}\hat{K}_\textsc{fbm}^{*}\right)^{\!-1\!}\!  1. \tag{B3}
\end{equation} 
Since the first term satisfies the normalization condition $(1, S)=1$ by itself, we conclude that $\lambda'\equiv 0$, and expression \eqref{eq:S_alpha>1_final} coincides with the result obtained in the main text for $\alpha<1$.

\textit{Case of ${\alpha=1}$.---}Since stationary $1/f$ noise is characterized by the logarithmic auto-correlation, 
we utilize the decomposition \cite{estrada2000singular}:
\begin{equation}\label{eq:Chebyshev_expansion_tau}
	-\ln\abs{\tau_1 - \tau_2} = 2\ln2 + \sum_{n=1}^{\infty} \frac{2}{n}\widetilde{T}_n(\tau_1) \widetilde{T}_n(\tau_2), \tag{B4}
\end{equation}
where $\widetilde{T}_n(\tau) \equiv T_n(2\tau -1)$ is the shifted Chebyshev polynomial of order $n$. These polynomials form a complete orthogonal basis on $[0,1]$ with the weight ${\left[\tau(1-\tau)\right]^{-1/2}}.$
Now, when \(\int_{0}^1 {\abs{S(\tau)}^2}\sqrt{\tau(1-\tau)}\dd\tau <\infty \) is satisfied, we can expand $S(\tau)$ into a series of Chebyshev polynomials:
\begin{equation}\label{eq:Chebyshev_expansion_S}
	S(\tau) = \frac{2}{\pi}\ \frac{\frac{1}{2}\Phi_0 + \Phi_1 \widetilde{T}_1(\tau) + \Phi_2 \widetilde{T}_2(\tau) + \ldots}{\sqrt{\tau(1-\tau)}}, \tag{B5}
\end{equation}
where the coefficients are found from $
\Phi_n = (\widetilde{T}_n, S)
$ for all $n$. Since $T_0(\tau) \equiv 1 $, we obtain $\Phi_0 = 1$ from the normalization condition $\left(1, S\right)=1$.
Using decompositions \eqref{eq:Chebyshev_expansion_tau} and \eqref{eq:Chebyshev_expansion_S}, we can now write $(S, \hat{R}S)\propto\left<\mathscr{F}\right>$ as a quadratic diagonal form of coefficients $\left\{\Phi_n\right\}$:
\begin{equation}\label{eq:quadratic_form_Chebyshev}
	\left(S, \hat{R}S\right) =
	4P_0\! \left[ -C - \ln(\frac{\pi f_{\min}T}{2}) +
	2\sum_{n=1}^\infty \frac{\Phi_n^2}{n} \right].\tag{B6}
\end{equation}
This expression reaches its minimum when $\Phi_n=0$ for all $ n\geq1$. Remarkably, the corresponding optimal shape $S_\text{opt}(\tau) =\tfrac{1}{\pi} \left[\tau(1-\tau)\right]^{-1/2}$ is a special case of formula \eqref{eq:S_alpha>1_final} with $\lambda'\equiv 0$ for $\alpha=1$.

This completes the proof that for the stationary $1/f^\alpha$ charge noise with any spectral exponent $\alpha\in(0,2)$, the optimal shape for a $\mathrm{SWAP}^k$ exchange pulse is given by the symmetric beta-distribution function with parameter $\left(1-\alpha/2\right)$. 

\section{Appendix C. Improved optimal pulse design procedure for $1/f^\alpha$ stationary noises}\label{appendix:autocor_approx}

The remainder term of order $O(z)$ in the series expansion of $E_\alpha(z)$ near $z\rightarrow0$ [Eq. \eqref{eq:exponential_integral_series}] is given by an infinite series $ \mathcal{R}(z) = \sum_{n=1}^{\infty} \frac{(-1)^{n-1} z^n }{n! (1-\alpha + n)}.$ Due to its oscillatory character and semblance to the Taylor expansion of $1-e^{-z}$, 
we approximate it as follows:
\begin{equation}\label{eq:remainder_exp_approximation}
	\mathcal{R}(z) \approx  A (1 - e^{-\zeta z}), \quad A = \frac{3-\alpha}{\left(2-\alpha\right)^2}, \quad \zeta = \frac{2-\alpha}{3-\alpha}.\tag{C1}
\end{equation}
This choice of parameters $A, \zeta$ ensures the accuracy of $E_\alpha(z)$ approximation up to the third order: $\mathcal{R}(z) - A (1 - e^{-\zeta z}) = O(z^3), \ z\rightarrow0.$ This yields an improved approximation for the stationary $1/f^\alpha$ noise auto-correlation, valid for all $\theta = 2\pi f_{\min} T \lesssim 1$: 
\begin{widetext}
\begin{equation}\label{eq:auto-correlation_improved_approx}
	R(\tau_1, \tau_2) \approx {4P_0}f_{\min}^{1-\alpha}\sin\frac{\pi \alpha}{2} \times
	\begin{cases}
		- C -\ln\theta-\ln\abs{\tau_1- \tau_2} + A\left(1-e^{-\theta\zeta \abs{\tau_1- \tau_2}}\right), & \alpha =1,	
		\\[10pt]
		\dfrac{1}{1-\alpha} \
		\qty\Big[  \Gamma({2-\alpha})\, \theta^{\alpha - 1}\abs{\tau_1- \tau_2}^{\alpha - 1} - 1]  + A\left(1-e^{-\theta\zeta \abs{\tau_1- \tau_2}}\right), &\alpha \neq 1.
	\end{cases}\tag{C2}
\end{equation}
\begin{figure}
\centering
\includegraphics[width=0.95\linewidth]{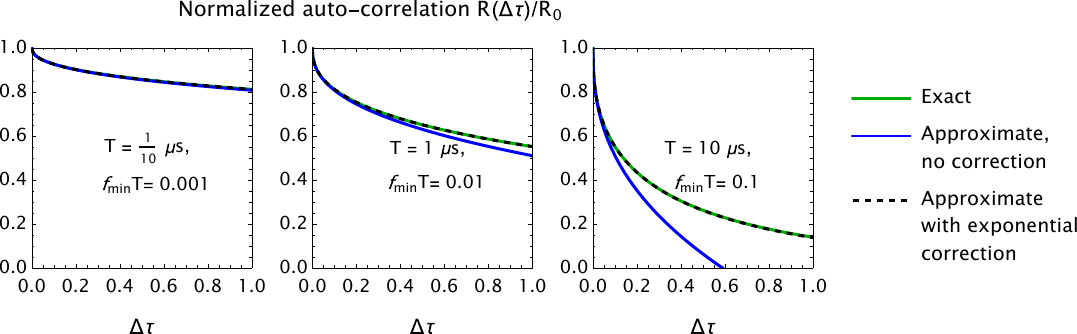}
\caption{Auto-correlation expressions for stationary $1/f^\alpha$ noise with $\alpha=1.4$: exact [Eq. \eqref{eq:auto-correlation_alpha}], approximated with the dominant term $\propto \left(\Delta\tau^{\alpha-1} +\text{const}\right)$ [Eq. \eqref{eq:auto-correlation_a<1}], and the approximation with exponential correction [Eq. \eqref{eq:auto-correlation_improved_approx}]. Here, $\Delta\tau = \abs{\tau_1 - \tau_2}$, $f_{\min}=10$ kHz, $f_{\max}=10$ GHz. }
\label{fig:autocorapprox}
\end{figure}
\end{widetext}
Figure \ref{fig:autocorapprox} compares the exact and approximate expressions for $R(\Delta \tau),$ where $\Delta\tau = \abs{\tau_1 - \tau_2}$. 
The coarse approximation (power-law for $\alpha\neq 1$ or logarithmic for $\alpha=1$) diverges from the exact expression most significantly in case of highly-correlated noises (large $\alpha$ values) and long pulses $T/T_{\exp}\sim 1.$ The improved approximation from Eq.  \eqref{eq:auto-correlation_improved_approx} shows good correspondence with the exact expression in all such cases. 

As follows from relation \eqref{eq:exponential_repr}, the exponential remainder term
\eqref{eq:remainder_exp_approximation} has a Caputo-Fabrizio fractional integral representation with $\tilde{\beta} = {1}/\left({\zeta\theta + 1}\right)$: 
\begin{equation*}\label{eq:remainder_CF}
	e^{-\theta \zeta \abs{\tau_1-\tau_2}} =  e^{-\theta\zeta \left(\tau_1+\tau_2\right)} + {2\tilde{\beta}^2\zeta\theta} \left({}_{0}\mathscr{I}^{\tilde{\beta}}\right)\! \left({}_{0}\mathscr{I}^{\tilde{\beta}}\right)^*\!\!\!\!.
\end{equation*}
This gives a precise approximation to the Lagrangian for $\mathrm{SWAP}^k$ exchange pulse optimization [Eq. \eqref{eq:Lagrangian}]. For example, for $\alpha<1$, we get:
\begin{align}\label{eq:Lagrangian_precise_R}
	\!\!\!\Lambda[S] \propto \text{const}+ \frac{\pi}{\sin\frac{\pi\alpha}{2}} \left(\hat{K}^* S ,\hat{K}^* S\right) - \lambda(1, S) \notag \\
 - A \left( e^{-\theta\zeta \tau}, S\right)^2 - {2\tilde{\beta}^2\zeta\theta} \left( \mathscr{I}_1^{\tilde{\beta}}S, \mathscr{I}_1^{\tilde{\beta}}S \right),\!\!\!\! \tag{C3}
\end{align}
where $\hat{K} =\tau^{\alpha/2}\, {}_{0}I^{\alpha/2} \tau^{-\alpha/2}$, introduced in the main text.
Although this functional is not positive definite, we previously proved the existence of an optimal exchange shape for a $\mathrm{SWAP}^k$ pulse for an arbitrary ensemble of TLFs [Eq. \eqref{eq:Lagrangian_general_dist}].
This suggests that the extremum condition $\delta\Lambda/\delta S =0$ will still give an accurate fractional integro-differential equation for the optimal shape:
\begin{multline}\label{eq:opt_shape_int_diff_eq}
	\frac{2\pi}{\sin\frac{\pi\alpha}{2}} \hat{K} \hat{K}^* S - \lambda - 2A \left( e^{-\theta\zeta \tau}, S\right) e^{-\theta\zeta \tau}  \\- {4\tilde{\beta}^2\zeta\theta}{}_0\mathscr{I}^{\tilde{\beta}} \mathscr{I}_1^{\tilde{\beta}}S=0.
 \tag{C4}
\end{multline}
As derived earlier, the zeroth-order approximation is a beta distribution function: $S_\text{opt}^{(0)}(\tau) \propto (\hat{K}  \hat{K}^*)^{-1} 1 \propto \tau^{-\alpha/2}  \left(1- \tau\right)^{-\alpha/2}.$
Therefore, by applying $\frac{1}{2\pi}{\sin\frac{\pi\alpha}{2}}(\hat{K}  \hat{K}^*)^{-1}$ to both sides of Eq. \eqref{eq:opt_shape_int_diff_eq}, we bring it to a form that can be solved by iterative methods such as fixed-point iteration:
 \begin{multline}\label{eq:opt_shape_int_diff_iterative}
 	S =\frac{ \tilde{\lambda} }{\left[\tau\left(1- \tau\right)\right]^{\frac{\alpha}{2}}} +  \frac{A\sin\frac{\pi\alpha}{2}\left( e^{-\theta\zeta \tau}, S\right)}{\pi}  (\hat{K}  \hat{K}^*)^{-1} e^{-\theta\zeta \tau}  \\+ 2 A\tilde{\beta}^2\zeta\theta \frac{\sin\frac{\pi\alpha}{2}}{\pi} (\hat{K}  \hat{K}^*)^{-1} {}_0\mathscr{I}^{\tilde{\beta}} \mathscr{I}_1^{\tilde{\beta}}S. \tag{C5}
 \end{multline} 
For the Lagrange multiplier $\tilde{\lambda}$, one would use the initial guess $\tilde{\lambda}^{(0)}= 1/{B\left(1-\frac{\alpha}{2}, 1-\frac{\alpha}{2}\right)},
$
and adjust it at each iteration step to ensure normalization $(1, S)=1$.
This procedure could be the most relevant to the long pulses and noise processes with large $\alpha$ values, where we expect the largest divergence from the beta-distribution shape from Eq. \eqref{eq:S_alpha>1_final} with $\lambda'\equiv0$.
However, we showed in the main text that the choice of a beta-distribution pulse shape gives nearly no gain in fidelity compared to a more experimentally feasible square shape. Thus, while finding an improved $S_\text{opt}(\tau)$ numerically is possible, it is not expected to noticeably change the infidelity compared to the beta pulse shape.  
 

\begin{thebibliography}{73}%
	\makeatletter
	\providecommand \@ifxundefined [1]{%
		\@ifx{#1\undefined}
	}%
	\providecommand \@ifnum [1]{%
		\ifnum #1\expandafter \@firstoftwo
		\else \expandafter \@secondoftwo
		\fi
	}%
	\providecommand \@ifx [1]{%
		\ifx #1\expandafter \@firstoftwo
		\else \expandafter \@secondoftwo
		\fi
	}%
	\providecommand \natexlab [1]{#1}%
	\providecommand \enquote  [1]{``#1''}%
	\providecommand \bibnamefont  [1]{#1}%
	\providecommand \bibfnamefont [1]{#1}%
	\providecommand \citenamefont [1]{#1}%
	\providecommand \href@noop [0]{\@secondoftwo}%
	\providecommand \href [0]{\begingroup \@sanitize@url \@href}%
	\providecommand \@href[1]{\@@startlink{#1}\@@href}%
	\providecommand \@@href[1]{\endgroup#1\@@endlink}%
	\providecommand \@sanitize@url [0]{\catcode `\\12\catcode `\$12\catcode
		`\&12\catcode `\#12\catcode `\^12\catcode `\_12\catcode `\%12\relax}%
	\providecommand \@@startlink[1]{}%
	\providecommand \@@endlink[0]{}%
	\providecommand \url  [0]{\begingroup\@sanitize@url \@url }%
	\providecommand \@url [1]{\endgroup\@href {#1}{\urlprefix }}%
	\providecommand \urlprefix  [0]{URL }%
	\providecommand \Eprint [0]{\href }%
	\providecommand \doibase [0]{https://doi.org/}%
	\providecommand \selectlanguage [0]{\@gobble}%
	\providecommand \bibinfo  [0]{\@secondoftwo}%
	\providecommand \bibfield  [0]{\@secondoftwo}%
	\providecommand \translation [1]{[#1]}%
	\providecommand \BibitemOpen [0]{}%
	\providecommand \bibitemStop [0]{}%
	\providecommand \bibitemNoStop [0]{.\EOS\space}%
	\providecommand \EOS [0]{\spacefactor3000\relax}%
	\providecommand \BibitemShut  [1]{\csname bibitem#1\endcsname}%
	\let\auto@bib@innerbib\@empty
	\bibitem [{\citenamefont {Culcer}\ \emph {et~al.}(2009)\citenamefont {Culcer},
		\citenamefont {Hu},\ and\ \citenamefont
		{Das~Sarma}}]{Culcer2009DephasingSispin}%
	\BibitemOpen
	\bibfield  {author} {\bibinfo {author} {\bibfnamefont {D.}~\bibnamefont
			{Culcer}}, \bibinfo {author} {\bibfnamefont {X.}~\bibnamefont {Hu}},\ and\
		\bibinfo {author} {\bibfnamefont {S.}~\bibnamefont {Das~Sarma}},\ }\bibfield
	{title} {\bibinfo {title} {Dephasing of {S}i spin qubits due to charge
			noise},\ }\bibfield  {journal} {\bibinfo  {journal} {Applied Physics
			Letters}\ }\textbf {\bibinfo {volume} {95}},\ \href
	{https://doi.org/10.1063/1.3194778} {10.1063/1.3194778} (\bibinfo {year}
	{2009})\BibitemShut {NoStop}%
	\bibitem [{\citenamefont {Burkard}\ \emph {et~al.}(2023)\citenamefont
		{Burkard}, \citenamefont {Ladd}, \citenamefont {Pan}, \citenamefont
		{Nichol},\ and\ \citenamefont {Petta}}]{Burkard2023Semiconductorspinqubits}%
	\BibitemOpen
	\bibfield  {author} {\bibinfo {author} {\bibfnamefont {G.}~\bibnamefont
			{Burkard}}, \bibinfo {author} {\bibfnamefont {T.~D.}\ \bibnamefont {Ladd}},
		\bibinfo {author} {\bibfnamefont {A.}~\bibnamefont {Pan}}, \bibinfo {author}
		{\bibfnamefont {J.~M.}\ \bibnamefont {Nichol}},\ and\ \bibinfo {author}
		{\bibfnamefont {J.~R.}\ \bibnamefont {Petta}},\ }\bibfield  {title} {\bibinfo
		{title} {Semiconductor spin qubits},\ }\href
	{https://doi.org/10.1103/revmodphys.95.025003} {\bibfield  {journal}
		{\bibinfo  {journal} {Reviews of Modern Physics}\ }\textbf {\bibinfo {volume}
			{95}},\ \bibinfo {pages} {025003} (\bibinfo {year} {2023})}\BibitemShut
	{NoStop}%
	\bibitem [{\citenamefont {Chan}\ \emph {et~al.}(2018)\citenamefont {Chan},
		\citenamefont {Huang}, \citenamefont {Yang}, \citenamefont {Hwang},
		\citenamefont {Hensen}, \citenamefont {Tanttu}, \citenamefont {Hudson},
		\citenamefont {Itoh}, \citenamefont {Laucht}, \citenamefont {Morello},\ and\
		\citenamefont {Dzurak}}]{Chan2018AssessmentEnvNoiseSpectroscopy}%
	\BibitemOpen
	\bibfield  {author} {\bibinfo {author} {\bibfnamefont {K.~W.}\ \bibnamefont
			{Chan}}, \bibinfo {author} {\bibfnamefont {W.}~\bibnamefont {Huang}},
		\bibinfo {author} {\bibfnamefont {C.~H.}\ \bibnamefont {Yang}}, \bibinfo
		{author} {\bibfnamefont {J.~C.~C.}\ \bibnamefont {Hwang}}, \bibinfo {author}
		{\bibfnamefont {B.}~\bibnamefont {Hensen}}, \bibinfo {author} {\bibfnamefont
			{T.}~\bibnamefont {Tanttu}}, \bibinfo {author} {\bibfnamefont {F.~E.}\
			\bibnamefont {Hudson}}, \bibinfo {author} {\bibfnamefont {K.~M.}\
			\bibnamefont {Itoh}}, \bibinfo {author} {\bibfnamefont {A.}~\bibnamefont
			{Laucht}}, \bibinfo {author} {\bibfnamefont {A.}~\bibnamefont {Morello}},\
		and\ \bibinfo {author} {\bibfnamefont {A.~S.}\ \bibnamefont {Dzurak}},\
	}\bibfield  {title} {\bibinfo {title} {Assessment of a silicon quantum dot
			spin qubit environment via noise spectroscopy},\ }\href
	{https://doi.org/10.1103/physrevapplied.10.044017} {\bibfield  {journal}
		{\bibinfo  {journal} {Physical Review Applied}\ }\textbf {\bibinfo {volume}
			{10}},\ \bibinfo {pages} {044017} (\bibinfo {year} {2018})}\BibitemShut
	{NoStop}%
	\bibitem [{\citenamefont {Connors}\ \emph {et~al.}(2019)\citenamefont
		{Connors}, \citenamefont {Nelson}, \citenamefont {Qiao}, \citenamefont
		{Edge},\ and\ \citenamefont {Nichol}}]{Connors2019NoiseSi/SiGe}%
	\BibitemOpen
	\bibfield  {author} {\bibinfo {author} {\bibfnamefont {E.~J.}\ \bibnamefont
			{Connors}}, \bibinfo {author} {\bibfnamefont {J.}~\bibnamefont {Nelson}},
		\bibinfo {author} {\bibfnamefont {H.}~\bibnamefont {Qiao}}, \bibinfo {author}
		{\bibfnamefont {L.~F.}\ \bibnamefont {Edge}},\ and\ \bibinfo {author}
		{\bibfnamefont {J.~M.}\ \bibnamefont {Nichol}},\ }\bibfield  {title}
	{\bibinfo {title} {Low-frequency charge noise in {S}i/{SiGe} quantum dots},\
	}\href {https://doi.org/10.1103/physrevb.100.165305} {\bibfield  {journal}
		{\bibinfo  {journal} {Physical Review B}\ }\textbf {\bibinfo {volume}
			{100}},\ \bibinfo {pages} {165305} (\bibinfo {year} {2019})}\BibitemShut
	{NoStop}%
	\bibitem [{\citenamefont {Kranz}\ \emph {et~al.}(2020)\citenamefont {Kranz},
		\citenamefont {Gorman}, \citenamefont {Thorgrimsson}, \citenamefont {He},
		\citenamefont {Keith}, \citenamefont {Keizer},\ and\ \citenamefont
		{Simmons}}]{Kranz2020SingleCrystalMinimizeNoise}%
	\BibitemOpen
	\bibfield  {author} {\bibinfo {author} {\bibfnamefont {L.}~\bibnamefont
			{Kranz}}, \bibinfo {author} {\bibfnamefont {S.~K.}\ \bibnamefont {Gorman}},
		\bibinfo {author} {\bibfnamefont {B.}~\bibnamefont {Thorgrimsson}}, \bibinfo
		{author} {\bibfnamefont {Y.}~\bibnamefont {He}}, \bibinfo {author}
		{\bibfnamefont {D.}~\bibnamefont {Keith}}, \bibinfo {author} {\bibfnamefont
			{J.~G.}\ \bibnamefont {Keizer}},\ and\ \bibinfo {author} {\bibfnamefont
			{M.~Y.}\ \bibnamefont {Simmons}},\ }\bibfield  {title} {\bibinfo {title}
		{Exploiting a single-crystal environment to minimize the charge noise on
			qubits in silicon},\ }\bibfield  {journal} {\bibinfo  {journal} {Advanced
			Materials}\ }\textbf {\bibinfo {volume} {32}},\ \href
	{https://doi.org/10.1002/adma.202003361} {10.1002/adma.202003361} (\bibinfo
	{year} {2020})\BibitemShut {NoStop}%
	\bibitem [{\citenamefont {Kuhlmann}\ \emph {et~al.}(2013)\citenamefont
		{Kuhlmann}, \citenamefont {Houel}, \citenamefont {Ludwig}, \citenamefont
		{Greuter}, \citenamefont {Reuter}, \citenamefont {Wieck}, \citenamefont
		{Poggio},\ and\ \citenamefont {Warburton}}]{Kuhlmann2013ChargeNoiseSpin}%
	\BibitemOpen
	\bibfield  {author} {\bibinfo {author} {\bibfnamefont {A.~V.}\ \bibnamefont
			{Kuhlmann}}, \bibinfo {author} {\bibfnamefont {J.}~\bibnamefont {Houel}},
		\bibinfo {author} {\bibfnamefont {A.}~\bibnamefont {Ludwig}}, \bibinfo
		{author} {\bibfnamefont {L.}~\bibnamefont {Greuter}}, \bibinfo {author}
		{\bibfnamefont {D.}~\bibnamefont {Reuter}}, \bibinfo {author} {\bibfnamefont
			{A.~D.}\ \bibnamefont {Wieck}}, \bibinfo {author} {\bibfnamefont
			{M.}~\bibnamefont {Poggio}},\ and\ \bibinfo {author} {\bibfnamefont {R.~J.}\
			\bibnamefont {Warburton}},\ }\bibfield  {title} {\bibinfo {title} {Charge
			noise and spin noise in a semiconductor quantum device},\ }\href
	{https://doi.org/10.1038/nphys2688} {\bibfield  {journal} {\bibinfo
			{journal} {Nature Physics}\ }\textbf {\bibinfo {volume} {9}},\ \bibinfo
		{pages} {570} (\bibinfo {year} {2013})}\BibitemShut {NoStop}%
	\bibitem [{\citenamefont {Yoneda}\ \emph {et~al.}(2017)\citenamefont {Yoneda},
		\citenamefont {Takeda}, \citenamefont {Otsuka}, \citenamefont {Nakajima},
		\citenamefont {Delbecq}, \citenamefont {Allison}, \citenamefont {Honda},
		\citenamefont {Kodera}, \citenamefont {Oda}, \citenamefont {Hoshi},
		\citenamefont {Usami}, \citenamefont {Itoh},\ and\ \citenamefont
		{Tarucha}}]{Yoneda2017quantumdotspin}%
	\BibitemOpen
	\bibfield  {author} {\bibinfo {author} {\bibfnamefont {J.}~\bibnamefont
			{Yoneda}}, \bibinfo {author} {\bibfnamefont {K.}~\bibnamefont {Takeda}},
		\bibinfo {author} {\bibfnamefont {T.}~\bibnamefont {Otsuka}}, \bibinfo
		{author} {\bibfnamefont {T.}~\bibnamefont {Nakajima}}, \bibinfo {author}
		{\bibfnamefont {M.~R.}\ \bibnamefont {Delbecq}}, \bibinfo {author}
		{\bibfnamefont {G.}~\bibnamefont {Allison}}, \bibinfo {author} {\bibfnamefont
			{T.}~\bibnamefont {Honda}}, \bibinfo {author} {\bibfnamefont
			{T.}~\bibnamefont {Kodera}}, \bibinfo {author} {\bibfnamefont
			{S.}~\bibnamefont {Oda}}, \bibinfo {author} {\bibfnamefont {Y.}~\bibnamefont
			{Hoshi}}, \bibinfo {author} {\bibfnamefont {N.}~\bibnamefont {Usami}},
		\bibinfo {author} {\bibfnamefont {K.~M.}\ \bibnamefont {Itoh}},\ and\
		\bibinfo {author} {\bibfnamefont {S.}~\bibnamefont {Tarucha}},\ }\bibfield
	{title} {\bibinfo {title} {A quantum-dot spin qubit with coherence limited by
			charge noise and fidelity higher than 99.9{\%}},\ }\href
	{https://doi.org/10.1038/s41565-017-0014-x} {\bibfield  {journal} {\bibinfo
			{journal} {Nature Nanotechnology}\ }\textbf {\bibinfo {volume} {13}},\
		\bibinfo {pages} {102} (\bibinfo {year} {2017})}\BibitemShut {NoStop}%
	\bibitem [{\citenamefont {Petit}\ \emph {et~al.}(2018)\citenamefont {Petit},
		\citenamefont {Boter}, \citenamefont {Eenink}, \citenamefont {Droulers},
		\citenamefont {Tagliaferri}, \citenamefont {Li}, \citenamefont {Franke},
		\citenamefont {Singh}, \citenamefont {Clarke}, \citenamefont {Schouten},
		\citenamefont {Dobrovitski}, \citenamefont {Vandersypen},\ and\ \citenamefont
		{Veldhorst}}]{Petit2018ChargeNoiseHotQubits}%
	\BibitemOpen
	\bibfield  {author} {\bibinfo {author} {\bibfnamefont {L.}~\bibnamefont
			{Petit}}, \bibinfo {author} {\bibfnamefont {J.}~\bibnamefont {Boter}},
		\bibinfo {author} {\bibfnamefont {H.}~\bibnamefont {Eenink}}, \bibinfo
		{author} {\bibfnamefont {G.}~\bibnamefont {Droulers}}, \bibinfo {author}
		{\bibfnamefont {M.}~\bibnamefont {Tagliaferri}}, \bibinfo {author}
		{\bibfnamefont {R.}~\bibnamefont {Li}}, \bibinfo {author} {\bibfnamefont
			{D.}~\bibnamefont {Franke}}, \bibinfo {author} {\bibfnamefont
			{K.}~\bibnamefont {Singh}}, \bibinfo {author} {\bibfnamefont
			{J.}~\bibnamefont {Clarke}}, \bibinfo {author} {\bibfnamefont
			{R.}~\bibnamefont {Schouten}}, \bibinfo {author} {\bibfnamefont
			{V.}~\bibnamefont {Dobrovitski}}, \bibinfo {author} {\bibfnamefont
			{L.}~\bibnamefont {Vandersypen}},\ and\ \bibinfo {author} {\bibfnamefont
			{M.}~\bibnamefont {Veldhorst}},\ }\bibfield  {title} {\bibinfo {title} {Spin
			lifetime and charge noise in hot silicon quantum dot qubits},\ }\href
	{https://doi.org/10.1103/physrevlett.121.076801} {\bibfield  {journal}
		{\bibinfo  {journal} {Physical Review Letters}\ }\textbf {\bibinfo {volume}
			{121}},\ \bibinfo {pages} {076801} (\bibinfo {year} {2018})}\BibitemShut
	{NoStop}%
	\bibitem [{\citenamefont {Struck}\ \emph {et~al.}(2020)\citenamefont {Struck},
		\citenamefont {Hollmann}, \citenamefont {Schauer}, \citenamefont {Fedorets},
		\citenamefont {Schmidbauer}, \citenamefont {Sawano}, \citenamefont {Riemann},
		\citenamefont {Abrosimov}, \citenamefont {Cywi{\'{n}}ski}, \citenamefont
		{Bougeard},\ and\ \citenamefont
		{Schreiber}}]{Struck2020EnergySplittingNoise_SiSiGe}%
	\BibitemOpen
	\bibfield  {author} {\bibinfo {author} {\bibfnamefont {T.}~\bibnamefont
			{Struck}}, \bibinfo {author} {\bibfnamefont {A.}~\bibnamefont {Hollmann}},
		\bibinfo {author} {\bibfnamefont {F.}~\bibnamefont {Schauer}}, \bibinfo
		{author} {\bibfnamefont {O.}~\bibnamefont {Fedorets}}, \bibinfo {author}
		{\bibfnamefont {A.}~\bibnamefont {Schmidbauer}}, \bibinfo {author}
		{\bibfnamefont {K.}~\bibnamefont {Sawano}}, \bibinfo {author} {\bibfnamefont
			{H.}~\bibnamefont {Riemann}}, \bibinfo {author} {\bibfnamefont {N.~V.}\
			\bibnamefont {Abrosimov}}, \bibinfo {author} {\bibfnamefont
			{{\L}.}~\bibnamefont {Cywi{\'{n}}ski}}, \bibinfo {author} {\bibfnamefont
			{D.}~\bibnamefont {Bougeard}},\ and\ \bibinfo {author} {\bibfnamefont
			{L.~R.}\ \bibnamefont {Schreiber}},\ }\bibfield  {title} {\bibinfo {title}
		{Low-frequency spin qubit energy splitting noise in highly purified
			${}^{28}${Si}/{SiGe}},\ }\bibfield  {journal} {\bibinfo  {journal} {npj
			Quantum Information}\ }\textbf {\bibinfo {volume} {6}},\ \href
	{https://doi.org/10.1038/s41534-020-0276-2} {10.1038/s41534-020-0276-2}
	(\bibinfo {year} {2020})\BibitemShut {NoStop}%
	\bibitem [{\citenamefont {Jock}\ \emph {et~al.}(2022)\citenamefont {Jock},
		\citenamefont {Jacobson}, \citenamefont {Rudolph}, \citenamefont {Ward},
		\citenamefont {Carroll},\ and\ \citenamefont
		{Luhman}}]{Jock2022SiliconSinglet–TripletQubit}%
	\BibitemOpen
	\bibfield  {author} {\bibinfo {author} {\bibfnamefont {R.~M.}\ \bibnamefont
			{Jock}}, \bibinfo {author} {\bibfnamefont {N.~T.}\ \bibnamefont {Jacobson}},
		\bibinfo {author} {\bibfnamefont {M.}~\bibnamefont {Rudolph}}, \bibinfo
		{author} {\bibfnamefont {D.~R.}\ \bibnamefont {Ward}}, \bibinfo {author}
		{\bibfnamefont {M.~S.}\ \bibnamefont {Carroll}},\ and\ \bibinfo {author}
		{\bibfnamefont {D.~R.}\ \bibnamefont {Luhman}},\ }\bibfield  {title}
	{\bibinfo {title} {A silicon singlet–triplet qubit driven by spin-valley
			coupling},\ }\bibfield  {journal} {\bibinfo  {journal} {Nature
			Communications}\ }\textbf {\bibinfo {volume} {13}},\ \href
	{https://doi.org/10.1038/s41467-022-28302-y} {10.1038/s41467-022-28302-y}
	(\bibinfo {year} {2022})\BibitemShut {NoStop}%
	\bibitem [{\citenamefont {Connors}\ \emph {et~al.}(2022)\citenamefont
		{Connors}, \citenamefont {Nelson}, \citenamefont {Edge},\ and\ \citenamefont
		{Nichol}}]{Connors2022ChargeNoiseSpectroscopy}%
	\BibitemOpen
	\bibfield  {author} {\bibinfo {author} {\bibfnamefont {E.~J.}\ \bibnamefont
			{Connors}}, \bibinfo {author} {\bibfnamefont {J.}~\bibnamefont {Nelson}},
		\bibinfo {author} {\bibfnamefont {L.~F.}\ \bibnamefont {Edge}},\ and\
		\bibinfo {author} {\bibfnamefont {J.~M.}\ \bibnamefont {Nichol}},\ }\bibfield
	{title} {\bibinfo {title} {Charge-noise spectroscopy of {Si}/{SiGe} quantum
			dots via dynamically-decoupled exchange oscillations},\ }\href
	{https://doi.org/10.1038/s41467-022-28519-x} {\bibfield  {journal} {\bibinfo
			{journal} {Nature Communications}\ }\textbf {\bibinfo {volume} {13}},\
		\bibinfo {pages} {940} (\bibinfo {year} {2022})}\BibitemShut {NoStop}%
	\bibitem [{\citenamefont {Lundberg}(2002)}]{Lundberg2002NoiseSourcesBulkCMOS}%
	\BibitemOpen
	\bibfield  {author} {\bibinfo {author} {\bibfnamefont {K.}~\bibnamefont
			{Lundberg}},\ }\bibfield  {title} {\bibinfo {title} {Noise sources in bulk
			{CMOS}}} (\bibinfo {year} {2002})\BibitemShut {NoStop}%
	\bibitem [{\citenamefont {de~Sousa}(2007)}]{Sousa2007Danglingbondspin}%
	\BibitemOpen
	\bibfield  {author} {\bibinfo {author} {\bibfnamefont {R.}~\bibnamefont
			{de~Sousa}},\ }\bibfield  {title} {\bibinfo {title} {Dangling-bond spin
			relaxation and magnetic $1/f$ noise from the amorphous-semiconductor/oxide
			interface: Theory},\ }\href {https://doi.org/10.1103/physrevb.76.245306}
	{\bibfield  {journal} {\bibinfo  {journal} {Physical Review B}\ }\textbf
		{\bibinfo {volume} {76}},\ \bibinfo {pages} {245306} (\bibinfo {year}
		{2007})}\BibitemShut {NoStop}%
	\bibitem [{\citenamefont {Kabytayev}\ \emph {et~al.}(2014)\citenamefont
		{Kabytayev}, \citenamefont {Green}, \citenamefont {Khodjasteh}, \citenamefont
		{Biercuk}, \citenamefont {Viola},\ and\ \citenamefont
		{Brown}}]{Kabytayev2014RobustnessCompositePulsesControlNoise}%
	\BibitemOpen
	\bibfield  {author} {\bibinfo {author} {\bibfnamefont {C.}~\bibnamefont
			{Kabytayev}}, \bibinfo {author} {\bibfnamefont {T.~J.}\ \bibnamefont
			{Green}}, \bibinfo {author} {\bibfnamefont {K.}~\bibnamefont {Khodjasteh}},
		\bibinfo {author} {\bibfnamefont {M.~J.}\ \bibnamefont {Biercuk}}, \bibinfo
		{author} {\bibfnamefont {L.}~\bibnamefont {Viola}},\ and\ \bibinfo {author}
		{\bibfnamefont {K.~R.}\ \bibnamefont {Brown}},\ }\bibfield  {title} {\bibinfo
		{title} {Robustness of composite pulses to time-dependent control noise},\
	}\href {https://doi.org/10.1103/PhysRevA.90.012316} {\bibfield  {journal}
		{\bibinfo  {journal} {Physical Review A}\ }\textbf {\bibinfo {volume} {90}},\
		\bibinfo {pages} {012316} (\bibinfo {year} {2014})}\BibitemShut {NoStop}%
	\bibitem [{\citenamefont {Shehata}\ \emph {et~al.}(2023)\citenamefont
		{Shehata}, \citenamefont {Simion}, \citenamefont {Li}, \citenamefont
		{Mohiyaddin}, \citenamefont {Wan}, \citenamefont {Mongillo}, \citenamefont
		{Govoreanu}, \citenamefont {Radu}, \citenamefont {De~Greve},\ and\
		\citenamefont {Van~Dorpe}}]{Shehata2023ModelingChargeEnvironment}%
	\BibitemOpen
	\bibfield  {author} {\bibinfo {author} {\bibfnamefont {M.~M. E.~K.}\
			\bibnamefont {Shehata}}, \bibinfo {author} {\bibfnamefont {G.}~\bibnamefont
			{Simion}}, \bibinfo {author} {\bibfnamefont {R.}~\bibnamefont {Li}}, \bibinfo
		{author} {\bibfnamefont {F.~A.}\ \bibnamefont {Mohiyaddin}}, \bibinfo
		{author} {\bibfnamefont {D.}~\bibnamefont {Wan}}, \bibinfo {author}
		{\bibfnamefont {M.}~\bibnamefont {Mongillo}}, \bibinfo {author}
		{\bibfnamefont {B.}~\bibnamefont {Govoreanu}}, \bibinfo {author}
		{\bibfnamefont {I.}~\bibnamefont {Radu}}, \bibinfo {author} {\bibfnamefont
			{K.}~\bibnamefont {De~Greve}},\ and\ \bibinfo {author} {\bibfnamefont
			{P.}~\bibnamefont {Van~Dorpe}},\ }\bibfield  {title} {\bibinfo {title}
		{Modeling semiconductor spin qubits and their charge noise environment for
			quantum gate fidelity estimation},\ }\href
	{https://doi.org/10.1103/physrevb.108.045305} {\bibfield  {journal} {\bibinfo
			{journal} {Physical Review B}\ }\textbf {\bibinfo {volume} {108}},\ \bibinfo
		{pages} {045305} (\bibinfo {year} {2023})}\BibitemShut {NoStop}%
	\bibitem [{\citenamefont {Paquelet~Wuetz}\ \emph {et~al.}(2023)\citenamefont
		{Paquelet~Wuetz}, \citenamefont {Degli~Esposti}, \citenamefont {Zwerver},
		\citenamefont {Amitonov}, \citenamefont {Botifoll}, \citenamefont {Arbiol},
		\citenamefont {Sammak}, \citenamefont {Vandersypen}, \citenamefont {Russ},\
		and\ \citenamefont {Scappucci}}]{PaqueletWuetz2023ReducingChargeNoise_QWs}%
	\BibitemOpen
	\bibfield  {author} {\bibinfo {author} {\bibfnamefont {B.}~\bibnamefont
			{Paquelet~Wuetz}}, \bibinfo {author} {\bibfnamefont {D.}~\bibnamefont
			{Degli~Esposti}}, \bibinfo {author} {\bibfnamefont {A.-M.~J.}\ \bibnamefont
			{Zwerver}}, \bibinfo {author} {\bibfnamefont {S.~V.}\ \bibnamefont
			{Amitonov}}, \bibinfo {author} {\bibfnamefont {M.}~\bibnamefont {Botifoll}},
		\bibinfo {author} {\bibfnamefont {J.}~\bibnamefont {Arbiol}}, \bibinfo
		{author} {\bibfnamefont {A.}~\bibnamefont {Sammak}}, \bibinfo {author}
		{\bibfnamefont {L.~M.~K.}\ \bibnamefont {Vandersypen}}, \bibinfo {author}
		{\bibfnamefont {M.}~\bibnamefont {Russ}},\ and\ \bibinfo {author}
		{\bibfnamefont {G.}~\bibnamefont {Scappucci}},\ }\bibfield  {title} {\bibinfo
		{title} {Reducing charge noise in quantum dots by using thin silicon quantum
			wells},\ }\href {https://doi.org/10.1038/s41467-023-36951-w} {\bibfield
		{journal} {\bibinfo  {journal} {Nature Communications}\ }\textbf {\bibinfo
			{volume} {14}},\ \bibinfo {pages} {1385} (\bibinfo {year}
		{2023})}\BibitemShut {NoStop}%
	\bibitem [{\citenamefont {Yang}\ \emph {et~al.}(2019)\citenamefont {Yang},
		\citenamefont {Chan}, \citenamefont {Harper}, \citenamefont {Huang},
		\citenamefont {Evans}, \citenamefont {Hwang}, \citenamefont {Hensen},
		\citenamefont {Laucht}, \citenamefont {Tanttu}, \citenamefont {Hudson},
		\citenamefont {Flammia}, \citenamefont {Itoh}, \citenamefont {Morello},
		\citenamefont {Bartlett},\ and\ \citenamefont
		{Dzurak}}]{Yang2019Siliconqubitfidelities}%
	\BibitemOpen
	\bibfield  {author} {\bibinfo {author} {\bibfnamefont {C.~H.}\ \bibnamefont
			{Yang}}, \bibinfo {author} {\bibfnamefont {K.~W.}\ \bibnamefont {Chan}},
		\bibinfo {author} {\bibfnamefont {R.}~\bibnamefont {Harper}}, \bibinfo
		{author} {\bibfnamefont {W.}~\bibnamefont {Huang}}, \bibinfo {author}
		{\bibfnamefont {T.}~\bibnamefont {Evans}}, \bibinfo {author} {\bibfnamefont
			{J.~C.~C.}\ \bibnamefont {Hwang}}, \bibinfo {author} {\bibfnamefont
			{B.}~\bibnamefont {Hensen}}, \bibinfo {author} {\bibfnamefont
			{A.}~\bibnamefont {Laucht}}, \bibinfo {author} {\bibfnamefont
			{T.}~\bibnamefont {Tanttu}}, \bibinfo {author} {\bibfnamefont {F.~E.}\
			\bibnamefont {Hudson}}, \bibinfo {author} {\bibfnamefont {S.~T.}\
			\bibnamefont {Flammia}}, \bibinfo {author} {\bibfnamefont {K.~M.}\
			\bibnamefont {Itoh}}, \bibinfo {author} {\bibfnamefont {A.}~\bibnamefont
			{Morello}}, \bibinfo {author} {\bibfnamefont {S.~D.}\ \bibnamefont
			{Bartlett}},\ and\ \bibinfo {author} {\bibfnamefont {A.~S.}\ \bibnamefont
			{Dzurak}},\ }\bibfield  {title} {\bibinfo {title} {Silicon qubit fidelities
			approaching incoherent noise limits via pulse engineering},\ }\href
	{https://doi.org/10.1038/s41928-019-0234-1} {\bibfield  {journal} {\bibinfo
			{journal} {Nature Electronics}\ }\textbf {\bibinfo {volume} {2}},\ \bibinfo
		{pages} {151} (\bibinfo {year} {2019})}\BibitemShut {NoStop}%
	\bibitem [{\citenamefont {Mills}\ \emph {et~al.}(2022)\citenamefont {Mills},
		\citenamefont {Guinn}, \citenamefont {Gullans}, \citenamefont {Sigillito},
		\citenamefont {Feldman}, \citenamefont {Nielsen},\ and\ \citenamefont
		{Petta}}]{Mills2022Twoqubitsilicon}%
	\BibitemOpen
	\bibfield  {author} {\bibinfo {author} {\bibfnamefont {A.~R.}\ \bibnamefont
			{Mills}}, \bibinfo {author} {\bibfnamefont {C.~R.}\ \bibnamefont {Guinn}},
		\bibinfo {author} {\bibfnamefont {M.~J.}\ \bibnamefont {Gullans}}, \bibinfo
		{author} {\bibfnamefont {A.~J.}\ \bibnamefont {Sigillito}}, \bibinfo {author}
		{\bibfnamefont {M.~M.}\ \bibnamefont {Feldman}}, \bibinfo {author}
		{\bibfnamefont {E.}~\bibnamefont {Nielsen}},\ and\ \bibinfo {author}
		{\bibfnamefont {J.~R.}\ \bibnamefont {Petta}},\ }\bibfield  {title} {\bibinfo
		{title} {Two-qubit silicon quantum processor with operation fidelity
			exceeding 99\%},\ }\bibfield  {journal} {\bibinfo  {journal} {Science
			Advances}\ }\textbf {\bibinfo {volume} {8}},\ \href
	{https://doi.org/10.1126/sciadv.abn5130} {10.1126/sciadv.abn5130} (\bibinfo
	{year} {2022})\BibitemShut {NoStop}%
	\bibitem [{\citenamefont {Rojas-Arias}\ \emph {et~al.}(2023)\citenamefont
		{Rojas-Arias}, \citenamefont {Noiri}, \citenamefont {Stano}, \citenamefont
		{Nakajima}, \citenamefont {Yoneda}, \citenamefont {Takeda}, \citenamefont
		{Kobayashi}, \citenamefont {Sammak}, \citenamefont {Scappucci}, \citenamefont
		{Loss},\ and\ \citenamefont
		{Tarucha}}]{RojasArias2023SpatialNoiseCorrelations}%
	\BibitemOpen
	\bibfield  {author} {\bibinfo {author} {\bibfnamefont {J.}~\bibnamefont
			{Rojas-Arias}}, \bibinfo {author} {\bibfnamefont {A.}~\bibnamefont {Noiri}},
		\bibinfo {author} {\bibfnamefont {P.}~\bibnamefont {Stano}}, \bibinfo
		{author} {\bibfnamefont {T.}~\bibnamefont {Nakajima}}, \bibinfo {author}
		{\bibfnamefont {J.}~\bibnamefont {Yoneda}}, \bibinfo {author} {\bibfnamefont
			{K.}~\bibnamefont {Takeda}}, \bibinfo {author} {\bibfnamefont
			{T.}~\bibnamefont {Kobayashi}}, \bibinfo {author} {\bibfnamefont
			{A.}~\bibnamefont {Sammak}}, \bibinfo {author} {\bibfnamefont
			{G.}~\bibnamefont {Scappucci}}, \bibinfo {author} {\bibfnamefont
			{D.}~\bibnamefont {Loss}},\ and\ \bibinfo {author} {\bibfnamefont
			{S.}~\bibnamefont {Tarucha}},\ }\bibfield  {title} {\bibinfo {title} {Spatial
			noise correlations beyond nearest neighbors in ${}^{28}${Si}/{Si}-{Ge} spin
			qubits},\ }\href {https://doi.org/10.1103/PhysRevApplied.20.054024}
	{\bibfield  {journal} {\bibinfo  {journal} {Physical Review Applied}\
		}\textbf {\bibinfo {volume} {20}},\ \bibinfo {pages} {054024} (\bibinfo
		{year} {2023})}\BibitemShut {NoStop}%
	\bibitem [{\citenamefont {Paladino}\ \emph {et~al.}(2014)\citenamefont
		{Paladino}, \citenamefont {Galperin}, \citenamefont {Falci},\ and\
		\citenamefont {Altshuler}}]{Paladino2014ChargeNoiseReview}%
	\BibitemOpen
	\bibfield  {author} {\bibinfo {author} {\bibfnamefont {E.}~\bibnamefont
			{Paladino}}, \bibinfo {author} {\bibfnamefont {Y.}~\bibnamefont {Galperin}},
		\bibinfo {author} {\bibfnamefont {G.}~\bibnamefont {Falci}},\ and\ \bibinfo
		{author} {\bibfnamefont {B.}~\bibnamefont {Altshuler}},\ }\bibfield  {title}
	{\bibinfo {title} {1/f noise: Implications for solid-state quantum
			information},\ }\href {https://doi.org/10.1103/revmodphys.86.361} {\bibfield
		{journal} {\bibinfo  {journal} {Reviews of Modern Physics}\ }\textbf
		{\bibinfo {volume} {86}},\ \bibinfo {pages} {361} (\bibinfo {year}
		{2014})}\BibitemShut {NoStop}%
	\bibitem [{\citenamefont {Viola}\ and\ \citenamefont
		{Lloyd}(1998)}]{Viola1998DynamicalSuppressionDecoherence}%
	\BibitemOpen
	\bibfield  {author} {\bibinfo {author} {\bibfnamefont {L.}~\bibnamefont
			{Viola}}\ and\ \bibinfo {author} {\bibfnamefont {S.}~\bibnamefont {Lloyd}},\
	}\bibfield  {title} {\bibinfo {title} {Dynamical suppression of decoherence
			in two-state quantum systems},\ }\href
	{https://doi.org/10.1103/PhysRevA.58.2733} {\bibfield  {journal} {\bibinfo
			{journal} {Physical Review A}\ }\textbf {\bibinfo {volume} {58}},\ \bibinfo
		{pages} {2733} (\bibinfo {year} {1998})}\BibitemShut {NoStop}%
	\bibitem [{\citenamefont {Shiokawa}\ and\ \citenamefont
		{Lidar}(2004)}]{Shiokawa2004DynamicaldecouplingSlowPulses}%
	\BibitemOpen
	\bibfield  {author} {\bibinfo {author} {\bibfnamefont {K.}~\bibnamefont
			{Shiokawa}}\ and\ \bibinfo {author} {\bibfnamefont {D.~A.}\ \bibnamefont
			{Lidar}},\ }\bibfield  {title} {\bibinfo {title} {Dynamical decoupling using
			slow pulses: {Efficient} suppression of $1/f$ noise},\ }\href
	{https://doi.org/10.1103/PhysRevA.69.030302} {\bibfield  {journal} {\bibinfo
			{journal} {Physical Review A}\ }\textbf {\bibinfo {volume} {69}},\ \bibinfo
		{pages} {030302} (\bibinfo {year} {2004})}\BibitemShut {NoStop}%
	\bibitem [{\citenamefont {Uhrig}(2007)}]{Uhrig2007KeepingQuantumBitAlive}%
	\BibitemOpen
	\bibfield  {author} {\bibinfo {author} {\bibfnamefont {G.~S.}\ \bibnamefont
			{Uhrig}},\ }\bibfield  {title} {\bibinfo {title} {Keeping a quantum bit alive
			by optimized $\pi$-{pulse} {sequences}},\ }\href
	{https://doi.org/10.1103/PhysRevLett.98.100504} {\bibfield  {journal}
		{\bibinfo  {journal} {Physical Review Letters}\ }\textbf {\bibinfo {volume}
			{98}},\ \bibinfo {pages} {100504} (\bibinfo {year} {2007})}\BibitemShut
	{NoStop}%
	\bibitem [{\citenamefont {Faoro}\ and\ \citenamefont
		{Viola}(2004)}]{Faoro2004DynamicalSuppression1/fnoise}%
	\BibitemOpen
	\bibfield  {author} {\bibinfo {author} {\bibfnamefont {L.}~\bibnamefont
			{Faoro}}\ and\ \bibinfo {author} {\bibfnamefont {L.}~\bibnamefont {Viola}},\
	}\bibfield  {title} {\bibinfo {title} {Dynamical suppression of 1/f noise
			processes in qubit systems},\ }\href
	{https://doi.org/10.1103/physrevlett.92.117905} {\bibfield  {journal}
		{\bibinfo  {journal} {Physical Review Letters}\ }\textbf {\bibinfo {volume}
			{92}},\ \bibinfo {pages} {117905} (\bibinfo {year} {2004})}\BibitemShut
	{NoStop}%
	\bibitem [{\citenamefont {Falci}\ \emph {et~al.}(2004)\citenamefont {Falci},
		\citenamefont {D’Arrigo}, \citenamefont {Mastellone},\ and\ \citenamefont
		{Paladino}}]{Falci2004DynamicalSuppressionTelegraphNoise}%
	\BibitemOpen
	\bibfield  {author} {\bibinfo {author} {\bibfnamefont {G.}~\bibnamefont
			{Falci}}, \bibinfo {author} {\bibfnamefont {A.}~\bibnamefont {D’Arrigo}},
		\bibinfo {author} {\bibfnamefont {A.}~\bibnamefont {Mastellone}},\ and\
		\bibinfo {author} {\bibfnamefont {E.}~\bibnamefont {Paladino}},\ }\bibfield
	{title} {\bibinfo {title} {Dynamical suppression of telegraph and 1/f noise
			due to quantum bistable fluctuators},\ }\href
	{https://doi.org/10.1103/physreva.70.040101} {\bibfield  {journal} {\bibinfo
			{journal} {Physical Review A}\ }\textbf {\bibinfo {volume} {70}},\ \bibinfo
		{pages} {040101} (\bibinfo {year} {2004})}\BibitemShut {NoStop}%
	\bibitem [{\citenamefont {Cywi\'{n}ski}\ \emph {et~al.}(2008)\citenamefont
		{Cywi\'{n}ski}, \citenamefont {Lutchyn}, \citenamefont {Nave},\ and\
		\citenamefont {Das~Sarma}}]{Cywinski2008HowEnhanceDephasing}%
	\BibitemOpen
	\bibfield  {author} {\bibinfo {author} {\bibfnamefont {{\L}.}~\bibnamefont
			{Cywi\'{n}ski}}, \bibinfo {author} {\bibfnamefont {R.~M.}\ \bibnamefont
			{Lutchyn}}, \bibinfo {author} {\bibfnamefont {C.~P.}\ \bibnamefont {Nave}},\
		and\ \bibinfo {author} {\bibfnamefont {S.}~\bibnamefont {Das~Sarma}},\
	}\bibfield  {title} {\bibinfo {title} {How to enhance dephasing time in
			superconducting qubits},\ }\href {https://doi.org/10.1103/physrevb.77.174509}
	{\bibfield  {journal} {\bibinfo  {journal} {Physical Review B}\ }\textbf
		{\bibinfo {volume} {77}},\ \bibinfo {pages} {174509} (\bibinfo {year}
		{2008})}\BibitemShut {NoStop}%
	\bibitem [{\citenamefont {Pasini}\ and\ \citenamefont
		{Uhrig}(2010)}]{Pasini2010DynamicalDecouplingPowerLawNoise}%
	\BibitemOpen
	\bibfield  {author} {\bibinfo {author} {\bibfnamefont {S.}~\bibnamefont
			{Pasini}}\ and\ \bibinfo {author} {\bibfnamefont {G.~S.}\ \bibnamefont
			{Uhrig}},\ }\bibfield  {title} {\bibinfo {title} {Optimized dynamical
			decoupling for power-law noise spectra},\ }\href
	{https://doi.org/10.1103/physreva.81.012309} {\bibfield  {journal} {\bibinfo
			{journal} {Physical Review A}\ }\textbf {\bibinfo {volume} {81}},\ \bibinfo
		{pages} {012309} (\bibinfo {year} {2010})}\BibitemShut {NoStop}%
	\bibitem [{\citenamefont {Ramon}(2015)}]{Ramon2015NonGaussiansignatures}%
	\BibitemOpen
	\bibfield  {author} {\bibinfo {author} {\bibfnamefont {G.}~\bibnamefont
			{Ramon}},\ }\bibfield  {title} {\bibinfo {title} {Non-{G}aussian signatures
			and collective effects in charge noise affecting a dynamically decoupled
			qubit},\ }\href {https://doi.org/10.1103/physrevb.92.155422} {\bibfield
		{journal} {\bibinfo  {journal} {Physical Review B}\ }\textbf {\bibinfo
			{volume} {92}},\ \bibinfo {pages} {155422} (\bibinfo {year}
		{2015})}\BibitemShut {NoStop}%
	\bibitem [{\citenamefont {Möttönen}\ \emph {et~al.}(2006)\citenamefont
		{Möttönen}, \citenamefont {de~Sousa}, \citenamefont {Zhang},\ and\
		\citenamefont {Whaley}}]{Moettoenen2006Highfidelity1QbitGatesChargeNoise}%
	\BibitemOpen
	\bibfield  {author} {\bibinfo {author} {\bibfnamefont {M.}~\bibnamefont
			{Möttönen}}, \bibinfo {author} {\bibfnamefont {R.}~\bibnamefont
			{de~Sousa}}, \bibinfo {author} {\bibfnamefont {J.}~\bibnamefont {Zhang}},\
		and\ \bibinfo {author} {\bibfnamefont {K.~B.}\ \bibnamefont {Whaley}},\
	}\bibfield  {title} {\bibinfo {title} {High-fidelity one-qubit operations
			under random telegraph noise},\ }\href
	{https://doi.org/10.1103/PhysRevA.73.022332} {\bibfield  {journal} {\bibinfo
			{journal} {Physical Review A}\ }\textbf {\bibinfo {volume} {73}},\ \bibinfo
		{pages} {022332} (\bibinfo {year} {2006})}\BibitemShut {NoStop}%
	\bibitem [{\citenamefont {Rebentrost}\ \emph {et~al.}(2009)\citenamefont
		{Rebentrost}, \citenamefont {Serban}, \citenamefont {Schulte-Herbrüggen},\
		and\ \citenamefont {Wilhelm}}]{Rebentrost2009OptimalControlNonMarkovianEnv}%
	\BibitemOpen
	\bibfield  {author} {\bibinfo {author} {\bibfnamefont {P.}~\bibnamefont
			{Rebentrost}}, \bibinfo {author} {\bibfnamefont {I.}~\bibnamefont {Serban}},
		\bibinfo {author} {\bibfnamefont {T.}~\bibnamefont {Schulte-Herbrüggen}},\
		and\ \bibinfo {author} {\bibfnamefont {F.~K.}\ \bibnamefont {Wilhelm}},\
	}\bibfield  {title} {\bibinfo {title} {Optimal control of a qubit coupled to
			a non-{Markovian} environment},\ }\href
	{https://doi.org/10.1103/PhysRevLett.102.090401} {\bibfield  {journal}
		{\bibinfo  {journal} {Physical Review Letters}\ }\textbf {\bibinfo {volume}
			{102}},\ \bibinfo {pages} {090401} (\bibinfo {year} {2009})}\BibitemShut
	{NoStop}%
	\bibitem [{\citenamefont {Pasini}\ \emph {et~al.}(2008)\citenamefont {Pasini},
		\citenamefont {Fischer}, \citenamefont {Karbach},\ and\ \citenamefont
		{Uhrig}}]{Pasini2008OptimizationShortCoherent}%
	\BibitemOpen
	\bibfield  {author} {\bibinfo {author} {\bibfnamefont {S.}~\bibnamefont
			{Pasini}}, \bibinfo {author} {\bibfnamefont {T.}~\bibnamefont {Fischer}},
		\bibinfo {author} {\bibfnamefont {P.}~\bibnamefont {Karbach}},\ and\ \bibinfo
		{author} {\bibfnamefont {G.~S.}\ \bibnamefont {Uhrig}},\ }\bibfield  {title}
	{\bibinfo {title} {Optimization of short coherent control pulses},\ }\href
	{https://doi.org/10.1103/PhysRevA.77.032315} {\bibfield  {journal} {\bibinfo
			{journal} {Physical Review A}\ }\textbf {\bibinfo {volume} {77}},\ \bibinfo
		{pages} {032315} (\bibinfo {year} {2008})}\BibitemShut {NoStop}%
	\bibitem [{\citenamefont {Kuopanportti}\ \emph {et~al.}(2008)\citenamefont
		{Kuopanportti}, \citenamefont {Möttönen}, \citenamefont {Bergholm},
		\citenamefont {Saira}, \citenamefont {Zhang},\ and\ \citenamefont
		{Whaley}}]{Kuopanportti2008Suppression1/fnoise}%
	\BibitemOpen
	\bibfield  {author} {\bibinfo {author} {\bibfnamefont {P.}~\bibnamefont
			{Kuopanportti}}, \bibinfo {author} {\bibfnamefont {M.}~\bibnamefont
			{Möttönen}}, \bibinfo {author} {\bibfnamefont {V.}~\bibnamefont
			{Bergholm}}, \bibinfo {author} {\bibfnamefont {O.-P.}\ \bibnamefont {Saira}},
		\bibinfo {author} {\bibfnamefont {J.}~\bibnamefont {Zhang}},\ and\ \bibinfo
		{author} {\bibfnamefont {K.~B.}\ \bibnamefont {Whaley}},\ }\bibfield  {title}
	{\bibinfo {title} {Suppression of $1/f^\alpha$ noise in one-qubit systems},\
	}\href {https://doi.org/10.1103/physreva.77.032334} {\bibfield  {journal}
		{\bibinfo  {journal} {Physical Review A}\ }\textbf {\bibinfo {volume} {77}},\
		\bibinfo {pages} {032334} (\bibinfo {year} {2008})}\BibitemShut {NoStop}%
	\bibitem [{\citenamefont {Zhang}\ \emph {et~al.}(2014)\citenamefont {Zhang},
		\citenamefont {Souza}, \citenamefont {Brandao},\ and\ \citenamefont
		{Suter}}]{Zhang2014ProtectedQuantumComputing}%
	\BibitemOpen
	\bibfield  {author} {\bibinfo {author} {\bibfnamefont {J.}~\bibnamefont
			{Zhang}}, \bibinfo {author} {\bibfnamefont {A.~M.}\ \bibnamefont {Souza}},
		\bibinfo {author} {\bibfnamefont {F.~D.}\ \bibnamefont {Brandao}},\ and\
		\bibinfo {author} {\bibfnamefont {D.}~\bibnamefont {Suter}},\ }\bibfield
	{title} {\bibinfo {title} {Protected quantum computing: interleaving gate
			operations with dynamical decoupling sequences},\ }\href
	{https://doi.org/10.1103/PhysRevLett.112.050502} {\bibfield  {journal}
		{\bibinfo  {journal} {Physical Review Letters}\ }\textbf {\bibinfo {volume}
			{112}},\ \bibinfo {pages} {050502} (\bibinfo {year} {2014})}\BibitemShut
	{NoStop}%
	\bibitem [{\citenamefont {D'Arrigo}\ \emph {et~al.}(2016)\citenamefont
		{D'Arrigo}, \citenamefont {Falci},\ and\ \citenamefont
		{Paladino}}]{DArrigo2016Highfidelity2qGatesDynamicalDecoupling}%
	\BibitemOpen
	\bibfield  {author} {\bibinfo {author} {\bibfnamefont {A.}~\bibnamefont
			{D'Arrigo}}, \bibinfo {author} {\bibfnamefont {G.}~\bibnamefont {Falci}},\
		and\ \bibinfo {author} {\bibfnamefont {E.}~\bibnamefont {Paladino}},\
	}\bibfield  {title} {\bibinfo {title} {High-fidelity two-qubit gates via
			dynamical decoupling of local $1/f$ noise at the optimal point},\ }\href
	{https://doi.org/10.1103/PhysRevA.94.022303} {\bibfield  {journal} {\bibinfo
			{journal} {Physical Review A}\ }\textbf {\bibinfo {volume} {94}},\ \bibinfo
		{pages} {022303} (\bibinfo {year} {2016})}\BibitemShut {NoStop}%
	\bibitem [{\citenamefont {Ram}\ \emph {et~al.}(2022)\citenamefont {Ram},
		\citenamefont {Krithika}, \citenamefont {Batra},\ and\ \citenamefont
		{Mahesh}}]{Ram2022RobustQuantumControl}%
	\BibitemOpen
	\bibfield  {author} {\bibinfo {author} {\bibfnamefont {M.~H.}\ \bibnamefont
			{Ram}}, \bibinfo {author} {\bibfnamefont {V.~R.}\ \bibnamefont {Krithika}},
		\bibinfo {author} {\bibfnamefont {P.}~\bibnamefont {Batra}},\ and\ \bibinfo
		{author} {\bibfnamefont {T.~S.}\ \bibnamefont {Mahesh}},\ }\bibfield  {title}
	{\bibinfo {title} {Robust quantum control using hybrid pulse engineering},\
	}\href {https://doi.org/10.1103/PhysRevA.105.042437} {\bibfield  {journal}
		{\bibinfo  {journal} {Physical Review A}\ }\textbf {\bibinfo {volume}
			{105}},\ \bibinfo {pages} {042437} (\bibinfo {year} {2022})}\BibitemShut
	{NoStop}%
	\bibitem [{\citenamefont {Zeng}\ \emph {et~al.}(2018)\citenamefont {Zeng},
		\citenamefont {Deng}, \citenamefont {Russo},\ and\ \citenamefont
		{Barnes}}]{Zeng2018GeneralSolutionInhomogeneousDephasing}%
	\BibitemOpen
	\bibfield  {author} {\bibinfo {author} {\bibfnamefont {J.}~\bibnamefont
			{Zeng}}, \bibinfo {author} {\bibfnamefont {X.-H.}\ \bibnamefont {Deng}},
		\bibinfo {author} {\bibfnamefont {A.}~\bibnamefont {Russo}},\ and\ \bibinfo
		{author} {\bibfnamefont {E.}~\bibnamefont {Barnes}},\ }\bibfield  {title}
	{\bibinfo {title} {General solution to inhomogeneous dephasing and smooth
			pulse dynamical decoupling},\ }\href
	{https://doi.org/10.1088/1367-2630/aaafe9} {\bibfield  {journal} {\bibinfo
			{journal} {New Journal of Physics}\ }\textbf {\bibinfo {volume} {20}},\
		\bibinfo {pages} {033011} (\bibinfo {year} {2018})}\BibitemShut {NoStop}%
	\bibitem [{\citenamefont {Zeng}\ and\ \citenamefont
		{Barnes}(2018)}]{Zeng2018FastestpulsesDynamicallyCorrected_PhaseGates}%
	\BibitemOpen
	\bibfield  {author} {\bibinfo {author} {\bibfnamefont {J.}~\bibnamefont
			{Zeng}}\ and\ \bibinfo {author} {\bibfnamefont {E.}~\bibnamefont {Barnes}},\
	}\bibfield  {title} {\bibinfo {title} {Fastest pulses that implement
			dynamically corrected single-qubit phase gates},\ }\href
	{https://doi.org/10.1103/physreva.98.012301} {\bibfield  {journal} {\bibinfo
			{journal} {Physical Review A}\ }\textbf {\bibinfo {volume} {98}},\ \bibinfo
		{pages} {012301} (\bibinfo {year} {2018})}\BibitemShut {NoStop}%
	\bibitem [{\citenamefont {Zeng}\ \emph {et~al.}(2019)\citenamefont {Zeng},
		\citenamefont {Yang}, \citenamefont {Dzurak},\ and\ \citenamefont
		{Barnes}}]{Zeng2019Geometricformalismconstructing}%
	\BibitemOpen
	\bibfield  {author} {\bibinfo {author} {\bibfnamefont {J.}~\bibnamefont
			{Zeng}}, \bibinfo {author} {\bibfnamefont {C.~H.}\ \bibnamefont {Yang}},
		\bibinfo {author} {\bibfnamefont {A.~S.}\ \bibnamefont {Dzurak}},\ and\
		\bibinfo {author} {\bibfnamefont {E.}~\bibnamefont {Barnes}},\ }\bibfield
	{title} {\bibinfo {title} {Geometric formalism for constructing arbitrary
			single-qubit dynamically corrected gates},\ }\href
	{https://doi.org/10.1103/physreva.99.052321} {\bibfield  {journal} {\bibinfo
			{journal} {Physical Review A}\ }\textbf {\bibinfo {volume} {99}},\ \bibinfo
		{pages} {052321} (\bibinfo {year} {2019})}\BibitemShut {NoStop}%
	\bibitem [{\citenamefont {Dong}\ \emph {et~al.}(2021)\citenamefont {Dong},
		\citenamefont {Zhuang}, \citenamefont {Economou},\ and\ \citenamefont
		{Barnes}}]{Dong2021DoublyGeometricQuantum}%
	\BibitemOpen
	\bibfield  {author} {\bibinfo {author} {\bibfnamefont {W.}~\bibnamefont
			{Dong}}, \bibinfo {author} {\bibfnamefont {F.}~\bibnamefont {Zhuang}},
		\bibinfo {author} {\bibfnamefont {S.~E.}\ \bibnamefont {Economou}},\ and\
		\bibinfo {author} {\bibfnamefont {E.}~\bibnamefont {Barnes}},\ }\bibfield
	{title} {\bibinfo {title} {Doubly geometric quantum control},\ }\href
	{https://doi.org/10.1103/prxquantum.2.030333} {\bibfield  {journal} {\bibinfo
			{journal} {{PRX} Quantum}\ }\textbf {\bibinfo {volume} {2}},\ \bibinfo
		{pages} {030333} (\bibinfo {year} {2021})}\BibitemShut {NoStop}%
	\bibitem [{\citenamefont {Nelson}\ \emph {et~al.}(2023)\citenamefont {Nelson},
		\citenamefont {Piliouras}, \citenamefont {Connelly},\ and\ \citenamefont
		{Barnes}}]{Nelson2023DynamicallyCorrectedGates_Multiple_Noise_Sources}%
	\BibitemOpen
	\bibfield  {author} {\bibinfo {author} {\bibfnamefont {H.~T.}\ \bibnamefont
			{Nelson}}, \bibinfo {author} {\bibfnamefont {E.}~\bibnamefont {Piliouras}},
		\bibinfo {author} {\bibfnamefont {K.}~\bibnamefont {Connelly}},\ and\
		\bibinfo {author} {\bibfnamefont {E.}~\bibnamefont {Barnes}},\ }\bibfield
	{title} {\bibinfo {title} {Designing dynamically corrected gates robust to
			multiple noise sources using geometric space curves},\ }\href
	{https://doi.org/10.1103/physreva.108.012407} {\bibfield  {journal} {\bibinfo
			{journal} {Physical Review A}\ }\textbf {\bibinfo {volume} {108}},\ \bibinfo
		{pages} {012407} (\bibinfo {year} {2023})}\BibitemShut {NoStop}%
	\bibitem [{\citenamefont {Buterakos}\ \emph {et~al.}(2021)\citenamefont
		{Buterakos}, \citenamefont {Sarma},\ and\ \citenamefont
		{Barnes}}]{Buterakos2021GeometricalFormalismDynamically}%
	\BibitemOpen
	\bibfield  {author} {\bibinfo {author} {\bibfnamefont {D.}~\bibnamefont
			{Buterakos}}, \bibinfo {author} {\bibfnamefont {S.~D.}\ \bibnamefont
			{Sarma}},\ and\ \bibinfo {author} {\bibfnamefont {E.}~\bibnamefont
			{Barnes}},\ }\bibfield  {title} {\bibinfo {title} {Geometrical formalism for
			dynamically corrected gates in multiqubit systems},\ }\href
	{https://doi.org/10.1103/prxquantum.2.010341} {\bibfield  {journal} {\bibinfo
			{journal} {{PRX} Quantum}\ }\textbf {\bibinfo {volume} {2}},\ \bibinfo
		{pages} {010341} (\bibinfo {year} {2021})}\BibitemShut {NoStop}%
	\bibitem [{\citenamefont {Samko}\ \emph {et~al.}(1993)\citenamefont {Samko},
		\citenamefont {Kilbas}, \citenamefont {Marichev} \emph
		{et~al.}}]{samko1993fractional}%
	\BibitemOpen
	\bibfield  {author} {\bibinfo {author} {\bibfnamefont {S.~G.}\ \bibnamefont
			{Samko}}, \bibinfo {author} {\bibfnamefont {A.~A.}\ \bibnamefont {Kilbas}},
		\bibinfo {author} {\bibfnamefont {O.~I.}\ \bibnamefont {Marichev}}, \emph
		{et~al.},\ }\href@noop {} {\emph {\bibinfo {title} {Fractional integrals and
				derivatives}}},\ Vol.~\bibinfo {volume} {1}\ (\bibinfo  {publisher} {Gordon
		and {B}reach science publishers, Yverdon-les-Bains, Switzerland},\ \bibinfo
	{year} {1993})\BibitemShut {NoStop}%
	\bibitem [{\citenamefont {Hu}\ and\ \citenamefont {Das~Sarma}(2006)}]{Hu_2006}%
	\BibitemOpen
	\bibfield  {author} {\bibinfo {author} {\bibfnamefont {X.}~\bibnamefont
			{Hu}}\ and\ \bibinfo {author} {\bibfnamefont {S.}~\bibnamefont {Das~Sarma}},\
	}\bibfield  {title} {\bibinfo {title} {Charge-fluctuation-induced dephasing
			of exchange-coupled spin qubits},\ }\href
	{https://doi.org/10.1103/physrevlett.96.100501} {\bibfield  {journal}
		{\bibinfo  {journal} {Physical Review Letters}\ }\textbf {\bibinfo {volume}
			{96}},\ \bibinfo {pages} {100501} (\bibinfo {year} {2006})}\BibitemShut
	{NoStop}%
	\bibitem [{\citenamefont {Burkard}\ \emph {et~al.}(1999)\citenamefont
		{Burkard}, \citenamefont {Loss},\ and\ \citenamefont
		{DiVincenzo}}]{Burkard_1999_architecture}%
	\BibitemOpen
	\bibfield  {author} {\bibinfo {author} {\bibfnamefont {G.}~\bibnamefont
			{Burkard}}, \bibinfo {author} {\bibfnamefont {D.}~\bibnamefont {Loss}},\ and\
		\bibinfo {author} {\bibfnamefont {D.~P.}\ \bibnamefont {DiVincenzo}},\
	}\bibfield  {title} {\bibinfo {title} {Coupled quantum dots as quantum
			gates},\ }\href {https://doi.org/10.1103/physrevb.59.2070} {\bibfield
		{journal} {\bibinfo  {journal} {Physical Review B}\ }\textbf {\bibinfo
			{volume} {59}},\ \bibinfo {pages} {2070} (\bibinfo {year}
		{1999})}\BibitemShut {NoStop}%
	\bibitem [{\citenamefont {Loss}\ and\ \citenamefont
		{DiVincenzo}(1998)}]{Loss_1998}%
	\BibitemOpen
	\bibfield  {author} {\bibinfo {author} {\bibfnamefont {D.}~\bibnamefont
			{Loss}}\ and\ \bibinfo {author} {\bibfnamefont {D.~P.}\ \bibnamefont
			{DiVincenzo}},\ }\bibfield  {title} {\bibinfo {title} {Quantum computation
			with quantum dots},\ }\href {https://doi.org/10.1103/physreva.57.120}
	{\bibfield  {journal} {\bibinfo  {journal} {Physical Review A}\ }\textbf
		{\bibinfo {volume} {57}},\ \bibinfo {pages} {120} (\bibinfo {year}
		{1998})}\BibitemShut {NoStop}%
	\bibitem [{\citenamefont {Stopa}\ and\ \citenamefont
		{Marcus}(2008)}]{Stopa2008MagneticFieldControl}%
	\BibitemOpen
	\bibfield  {author} {\bibinfo {author} {\bibfnamefont {M.}~\bibnamefont
			{Stopa}}\ and\ \bibinfo {author} {\bibfnamefont {C.~M.}\ \bibnamefont
			{Marcus}},\ }\bibfield  {title} {\bibinfo {title} {Magnetic {Field} {Control}
			of {Exchange} and {Noise} {Immunity} in {Double} {Quantum} {Dots}},\ }\href
	{https://doi.org/10.1021/nl801282t} {\bibfield  {journal} {\bibinfo
			{journal} {Nano Letters}\ }\textbf {\bibinfo {volume} {8}},\ \bibinfo {pages}
		{1778} (\bibinfo {year} {2008})}\BibitemShut {NoStop}%
	\bibitem [{\citenamefont {Shim}\ and\ \citenamefont
		{Tahan}(2016)}]{Shim2016ChargeNoiseInsensitive_ExchOnly}%
	\BibitemOpen
	\bibfield  {author} {\bibinfo {author} {\bibfnamefont {Y.-P.}\ \bibnamefont
			{Shim}}\ and\ \bibinfo {author} {\bibfnamefont {C.}~\bibnamefont {Tahan}},\
	}\bibfield  {title} {\bibinfo {title} {Charge-noise-insensitive gate
			operations for always-on, exchange-only qubits},\ }\href
	{https://doi.org/10.1103/PhysRevB.93.121410} {\bibfield  {journal} {\bibinfo
			{journal} {Physical Review B}\ }\textbf {\bibinfo {volume} {93}},\ \bibinfo
		{pages} {121410} (\bibinfo {year} {2016})}\BibitemShut {NoStop}%
	\bibitem [{\citenamefont {Reed}\ \emph {et~al.}(2016)\citenamefont {Reed},
		\citenamefont {Maune}, \citenamefont {Andrews}, \citenamefont {Borselli},
		\citenamefont {Eng}, \citenamefont {Jura}, \citenamefont {Kiselev},
		\citenamefont {Ladd}, \citenamefont {Merkel},\ and\ \citenamefont
		{et~al.}}]{Reed2016ReducedSensitivityChargeNoiseSymmetric}%
	\BibitemOpen
	\bibfield  {author} {\bibinfo {author} {\bibfnamefont {M.}~\bibnamefont
			{Reed}}, \bibinfo {author} {\bibfnamefont {B.}~\bibnamefont {Maune}},
		\bibinfo {author} {\bibfnamefont {R.}~\bibnamefont {Andrews}}, \bibinfo
		{author} {\bibfnamefont {M.}~\bibnamefont {Borselli}}, \bibinfo {author}
		{\bibfnamefont {K.}~\bibnamefont {Eng}}, \bibinfo {author} {\bibfnamefont
			{M.}~\bibnamefont {Jura}}, \bibinfo {author} {\bibfnamefont {A.}~\bibnamefont
			{Kiselev}}, \bibinfo {author} {\bibfnamefont {T.}~\bibnamefont {Ladd}},
		\bibinfo {author} {\bibfnamefont {S.}~\bibnamefont {Merkel}},\ and\ \bibinfo
		{author} {\bibfnamefont {I.~M.}\ \bibnamefont {et~al.}},\ }\bibfield  {title}
	{\bibinfo {title} {Reduced sensitivity to charge noise in semiconductor spin
			qubits via symmetric operation},\ }\href
	{https://doi.org/10.1103/physrevlett.116.110402} {\bibfield  {journal}
		{\bibinfo  {journal} {Physical Review Letters}\ }\textbf {\bibinfo {volume}
			{116}},\ \bibinfo {pages} {110402} (\bibinfo {year} {2016})}\BibitemShut
	{NoStop}%
	\bibitem [{\citenamefont {Martins}\ \emph {et~al.}(2016)\citenamefont
		{Martins}, \citenamefont {Malinowski}, \citenamefont {Nissen}, \citenamefont
		{Barnes}, \citenamefont {Fallahi}, \citenamefont {Gardner}, \citenamefont
		{Manfra}, \citenamefont {Marcus},\ and\ \citenamefont
		{Kuemmeth}}]{Martins2016NoiseSuppressionSymmetricExchangeGates}%
	\BibitemOpen
	\bibfield  {author} {\bibinfo {author} {\bibfnamefont {F.}~\bibnamefont
			{Martins}}, \bibinfo {author} {\bibfnamefont {F.~K.}\ \bibnamefont
			{Malinowski}}, \bibinfo {author} {\bibfnamefont {P.~D.}\ \bibnamefont
			{Nissen}}, \bibinfo {author} {\bibfnamefont {E.}~\bibnamefont {Barnes}},
		\bibinfo {author} {\bibfnamefont {S.}~\bibnamefont {Fallahi}}, \bibinfo
		{author} {\bibfnamefont {G.~C.}\ \bibnamefont {Gardner}}, \bibinfo {author}
		{\bibfnamefont {M.~J.}\ \bibnamefont {Manfra}}, \bibinfo {author}
		{\bibfnamefont {C.~M.}\ \bibnamefont {Marcus}},\ and\ \bibinfo {author}
		{\bibfnamefont {F.}~\bibnamefont {Kuemmeth}},\ }\bibfield  {title} {\bibinfo
		{title} {Noise suppression using symmetric exchange gates in spin qubits},\
	}\href {https://doi.org/10.1103/physrevlett.116.116801} {\bibfield  {journal}
		{\bibinfo  {journal} {Physical Review Letters}\ }\textbf {\bibinfo {volume}
			{116}},\ \bibinfo {pages} {116801} (\bibinfo {year} {2016})}\BibitemShut
	{NoStop}%
	\bibitem [{\citenamefont {Nielsen}\ \emph {et~al.}(2010)\citenamefont
		{Nielsen}, \citenamefont {Young}, \citenamefont {Muller},\ and\ \citenamefont
		{Carroll}}]{Nielsen2010Implicationssimultaneousrequirements}%
	\BibitemOpen
	\bibfield  {author} {\bibinfo {author} {\bibfnamefont {E.}~\bibnamefont
			{Nielsen}}, \bibinfo {author} {\bibfnamefont {R.~W.}\ \bibnamefont {Young}},
		\bibinfo {author} {\bibfnamefont {R.~P.}\ \bibnamefont {Muller}},\ and\
		\bibinfo {author} {\bibfnamefont {M.~S.}\ \bibnamefont {Carroll}},\
	}\bibfield  {title} {\bibinfo {title} {Implications of simultaneous
			requirements for low-noise exchange gates in double quantum dots},\ }\href
	{https://doi.org/10.1103/physrevb.82.075319} {\bibfield  {journal} {\bibinfo
			{journal} {Physical Review B}\ }\textbf {\bibinfo {volume} {82}},\ \bibinfo
		{pages} {075319} (\bibinfo {year} {2010})}\BibitemShut {NoStop}%
	\bibitem [{\citenamefont {Buonacorsi}(2021)}]{Buonacorsi2021Quantumdotdevices}%
	\BibitemOpen
	\bibfield  {author} {\bibinfo {author} {\bibfnamefont {B.}~\bibnamefont
			{Buonacorsi}},\ }\emph {\bibinfo {title} {Quantum dot devices in silicon and
			dopant-free GaAs/AlGaAs heterostructures}},\ \href@noop {} {Ph.D. thesis},\
	\bibinfo  {school} {University of Waterloo} (\bibinfo {year}
	{2021})\BibitemShut {NoStop}%
	\bibitem [{\citenamefont {Dial}\ \emph {et~al.}(2013)\citenamefont {Dial},
		\citenamefont {Shulman}, \citenamefont {Harvey}, \citenamefont {Bluhm},
		\citenamefont {Umansky},\ and\ \citenamefont
		{Yacoby}}]{Dial2013ChargeNoiseSpectroscopyExchOscillations}%
	\BibitemOpen
	\bibfield  {author} {\bibinfo {author} {\bibfnamefont {O.~E.}\ \bibnamefont
			{Dial}}, \bibinfo {author} {\bibfnamefont {M.~D.}\ \bibnamefont {Shulman}},
		\bibinfo {author} {\bibfnamefont {S.~P.}\ \bibnamefont {Harvey}}, \bibinfo
		{author} {\bibfnamefont {H.}~\bibnamefont {Bluhm}}, \bibinfo {author}
		{\bibfnamefont {V.}~\bibnamefont {Umansky}},\ and\ \bibinfo {author}
		{\bibfnamefont {A.}~\bibnamefont {Yacoby}},\ }\bibfield  {title} {\bibinfo
		{title} {Charge noise spectroscopy using coherent exchange oscillations in a
			singlet-triplet qubit},\ }\href
	{https://doi.org/10.1103/physrevlett.110.146804} {\bibfield  {journal}
		{\bibinfo  {journal} {Physical Review Letters}\ }\textbf {\bibinfo {volume}
			{110}},\ \bibinfo {pages} {146804} (\bibinfo {year} {2013})}\BibitemShut
	{NoStop}%
	\bibitem [{\citenamefont {Petta}\ \emph {et~al.}(2005)\citenamefont {Petta},
		\citenamefont {Johnson}, \citenamefont {Taylor}, \citenamefont {Laird},
		\citenamefont {Yacoby}, \citenamefont {Lukin}, \citenamefont {Marcus},
		\citenamefont {Hanson},\ and\ \citenamefont {Gossard}}]{Petta_2005}%
	\BibitemOpen
	\bibfield  {author} {\bibinfo {author} {\bibfnamefont {J.~R.}\ \bibnamefont
			{Petta}}, \bibinfo {author} {\bibfnamefont {A.~C.}\ \bibnamefont {Johnson}},
		\bibinfo {author} {\bibfnamefont {J.~M.}\ \bibnamefont {Taylor}}, \bibinfo
		{author} {\bibfnamefont {E.~A.}\ \bibnamefont {Laird}}, \bibinfo {author}
		{\bibfnamefont {A.}~\bibnamefont {Yacoby}}, \bibinfo {author} {\bibfnamefont
			{M.~D.}\ \bibnamefont {Lukin}}, \bibinfo {author} {\bibfnamefont {C.~M.}\
			\bibnamefont {Marcus}}, \bibinfo {author} {\bibfnamefont {M.~P.}\
			\bibnamefont {Hanson}},\ and\ \bibinfo {author} {\bibfnamefont {A.~C.}\
			\bibnamefont {Gossard}},\ }\bibfield  {title} {\bibinfo {title} {Coherent
			manipulation of coupled electron spins in semiconductor quantum dots},\
	}\href {https://doi.org/10.1126/science.1116955} {\bibfield  {journal}
		{\bibinfo  {journal} {Science}\ }\textbf {\bibinfo {volume} {309}},\ \bibinfo
		{pages} {2180} (\bibinfo {year} {2005})}\BibitemShut {NoStop}%
	\bibitem [{Note1()}]{Note1}%
	\BibitemOpen
	\bibinfo {note} {Note that according to Markov's inequality, average
		infidelity $\protect \langle \protect \mathscr {F}\protect \rangle $ also
		gives the upper bound on the extent of the tails of the distribution
		$\protect \mathscr {F}(\protect \widetilde {v})$}\BibitemShut {NoStop}%
	\bibitem [{\citenamefont
		{Yamamoto}(2017)}]{Yamamoto2017FundamentalsNoiseProcesses}%
	\BibitemOpen
	\bibfield  {author} {\bibinfo {author} {\bibfnamefont {Y.}~\bibnamefont
			{Yamamoto}},\ }\href
	{https://www.nii.ac.jp/qis/first-quantum/e/forStudents/lecture/index.html}
	{\emph {\bibinfo {title} {Fundamentals of Noise Processes}}}\ (\bibinfo
	{publisher} {University of Cambridge ESOL Examinations},\ \bibinfo {year}
	{2017})\BibitemShut {NoStop}%
	\bibitem [{\citenamefont {Milotti}(2002)}]{Milotti2002_1/fNoisePedagogical}%
	\BibitemOpen
	\bibfield  {author} {\bibinfo {author} {\bibfnamefont {E.}~\bibnamefont
			{Milotti}},\ }\bibfield  {title} {\bibinfo {title} {1/f noise: a pedagogical
			review},\ }\Eprint {https://arxiv.org/abs/physics/0204033}
	{arXiv:physics/0204033 [physics.class-ph]}  (\bibinfo {year}
	{2002})\BibitemShut {NoStop}%
	\bibitem [{\citenamefont {Astafiev}\ \emph {et~al.}(2004)\citenamefont
		{Astafiev}, \citenamefont {Pashkin}, \citenamefont {Nakamura}, \citenamefont
		{Yamamoto},\ and\ \citenamefont {Tsai}}]{Astafiev2004QuantumNoiseJosephson}%
	\BibitemOpen
	\bibfield  {author} {\bibinfo {author} {\bibfnamefont {O.}~\bibnamefont
			{Astafiev}}, \bibinfo {author} {\bibfnamefont {Y.~A.}\ \bibnamefont
			{Pashkin}}, \bibinfo {author} {\bibfnamefont {Y.}~\bibnamefont {Nakamura}},
		\bibinfo {author} {\bibfnamefont {T.}~\bibnamefont {Yamamoto}},\ and\
		\bibinfo {author} {\bibfnamefont {J.~S.}\ \bibnamefont {Tsai}},\ }\bibfield
	{title} {\bibinfo {title} {Quantum {Noise} in the {Josephson} {Charge}
			{Qubit}},\ }\href {https://doi.org/10.1103/PhysRevLett.93.267007} {\bibfield
		{journal} {\bibinfo  {journal} {Physical Review Letters}\ }\textbf {\bibinfo
			{volume} {93}},\ \bibinfo {pages} {267007} (\bibinfo {year}
		{2004})}\BibitemShut {NoStop}%
	\bibitem [{{\relax DLMF}()}]{NistDLMF}%
	\BibitemOpen
	{\relax DLMF},\ \href {https://dlmf.nist.gov/} {\bibinfo {title} {{\it NIST
				Digital Library of Mathematical Functions}}},\ \bibinfo {howpublished}
	{\url{https://dlmf.nist.gov/}, Release 1.2.0 of 2024-03-15},\ \bibinfo {note}
	{f.~W.~J. Olver, A.~B. {Olde Daalhuis}, D.~W. Lozier, B.~I. Schneider, R.~F.
		Boisvert, C.~W. Clark, B.~R. Miller, B.~V. Saunders, H.~S. Cohl, and M.~A.
		McClain, eds.}\BibitemShut {Stop}%
\BibitemShut
	{NoStop}%
	\bibitem [{\citenamefont
		{Anastassiou}(2009)}]{Anastassiou2009FractionalDifferentiationInequalities}%
	\BibitemOpen
	\bibfield  {author} {\bibinfo {author} {\bibfnamefont {G.~A.}\ \bibnamefont
			{Anastassiou}},\ }\href@noop {} {\emph {\bibinfo {title} {Fractional
				Differentiation Inequalities}}}\ (\bibinfo  {publisher} {Springer-Verlag New
		York},\ \bibinfo {year} {2009})\BibitemShut {NoStop}%
	\bibitem [{\citenamefont {Abdeljawad}\ \emph {et~al.}(2019)\citenamefont
		{Abdeljawad}, \citenamefont {Atangana}, \citenamefont {G{\'o}mez-Aguilar},\
		and\ \citenamefont {Jarad}}]{abdeljawad2019by_parts}%
	\BibitemOpen
	\bibfield  {author} {\bibinfo {author} {\bibfnamefont {T.}~\bibnamefont
			{Abdeljawad}}, \bibinfo {author} {\bibfnamefont {A.}~\bibnamefont
			{Atangana}}, \bibinfo {author} {\bibfnamefont {J.}~\bibnamefont
			{G{\'o}mez-Aguilar}},\ and\ \bibinfo {author} {\bibfnamefont
			{F.}~\bibnamefont {Jarad}},\ }\bibfield  {title} {\bibinfo {title} {On a more
			general fractional integration by parts formulae and applications},\
	}\href@noop {} {\bibfield  {journal} {\bibinfo  {journal} {Physica A:
				Statistical Mechanics and its Applications}\ }\textbf {\bibinfo {volume}
			{536}},\ \bibinfo {pages} {122494} (\bibinfo {year} {2019})}\BibitemShut
	{NoStop}%
	\bibitem [{Note2()}]{Note2}%
	\BibitemOpen
	\bibinfo {note} {The conjugation relation \protect \ref {eq:int_conjugate}
		for $\protect \beta >0$ holds for $\protect \varphi (x) \in L\protect ^p(a,
		b)$, $\protect \psi (x) \protect \in L\protect ^q(a, b)$ satisfying $\ p, q
		\protect \geq 1$, and ${p}^{-1} + {q}^{-1} \protect \leq 1 + \protect \beta
		$; in case of equality, $p, q \protect \neq 1$.}\BibitemShut {Stop}%
	\bibitem [{\citenamefont {Williams}(1963)}]{williams1963class_int_eq}%
	\BibitemOpen
	\bibfield  {author} {\bibinfo {author} {\bibfnamefont {W.~E.}\ \bibnamefont
			{Williams}},\ }\bibfield  {title} {\bibinfo {title} {A class of integral
			equations},\ }\bibfield  {booktitle} {\emph {\bibinfo {booktitle}
			{Mathematical Proceedings of the Cambridge Philosophical Society}},\ }\href
	{https://doi.org/10.1017/s0305004100037269} {\bibfield  {journal} {\bibinfo
			{journal} {Mathematical Proceedings of the Cambridge Philosophical Society}\
		}\textbf {\bibinfo {volume} {59}},\ \bibinfo {pages} {589} (\bibinfo {year}
		{1963})}\BibitemShut {NoStop}%
	\bibitem [{\citenamefont {Decreusefond}\ and\ \citenamefont
		{\"{U}st\"{u}nel}(1999)}]{Decreusefond1999StochasticAnalysisFBM}%
	\BibitemOpen
	\bibfield  {author} {\bibinfo {author} {\bibfnamefont {L.}~\bibnamefont
			{Decreusefond}}\ and\ \bibinfo {author} {\bibfnamefont {A.}~\bibnamefont
			{\"{U}st\"{u}nel}},\ }\bibfield  {title} {\bibinfo {title} {Stochastic
			analysis of the fractional {B}rownian motion},\ }\href
	{https://doi.org/10.1023/a:1008634027843} {\bibfield  {journal} {\bibinfo
			{journal} {Potential Analysis}\ }\textbf {\bibinfo {volume} {10}},\ \bibinfo
		{pages} {177} (\bibinfo {year} {1999})}\BibitemShut {NoStop}%
	\bibitem [{\citenamefont {Buonacorsi}\ \emph {et~al.}(2020)\citenamefont
		{Buonacorsi}, \citenamefont {Korkusinski}, \citenamefont {Khromets},\ and\
		\citenamefont {Baugh}}]{Buonacorsi2020Optimizinglateralquantum}%
	\BibitemOpen
	\bibfield  {author} {\bibinfo {author} {\bibfnamefont {B.}~\bibnamefont
			{Buonacorsi}}, \bibinfo {author} {\bibfnamefont {M.}~\bibnamefont
			{Korkusinski}}, \bibinfo {author} {\bibfnamefont {B.}~\bibnamefont
			{Khromets}},\ and\ \bibinfo {author} {\bibfnamefont {J.}~\bibnamefont
			{Baugh}},\ }\bibfield  {title} {\bibinfo {title} {Optimizing lateral quantum
			dot geometries for reduced exchange noise},\ }\Eprint
	{https://arxiv.org/abs/2012.10512} {arXiv:2012.10512 [cond-mat.mes-hall]}
	(\bibinfo {year} {2020})\BibitemShut {NoStop}%
	\bibitem [{\citenamefont {Mandelbrot}\ and\ \citenamefont
		{Van~Ness}(1968)}]{mandelbrot1968fractional}%
	\BibitemOpen
	\bibfield  {author} {\bibinfo {author} {\bibfnamefont {B.~B.}\ \bibnamefont
			{Mandelbrot}}\ and\ \bibinfo {author} {\bibfnamefont {J.~W.}\ \bibnamefont
			{Van~Ness}},\ }\bibfield  {title} {\bibinfo {title} {Fractional brownian
			motions, fractional noises and applications},\ }\href@noop {} {\bibfield
		{journal} {\bibinfo  {journal} {SIAM review}\ }\textbf {\bibinfo {volume}
			{10}},\ \bibinfo {pages} {422} (\bibinfo {year} {1968})}\BibitemShut
	{NoStop}%
	\bibitem [{\citenamefont {Flandrin}(1989)}]{flandrin1989spectrum}%
	\BibitemOpen
	\bibfield  {author} {\bibinfo {author} {\bibfnamefont {P.}~\bibnamefont
			{Flandrin}},\ }\bibfield  {title} {\bibinfo {title} {On the spectrum of
			fractional {B}rownian motions},\ }\href@noop {} {\bibfield  {journal}
		{\bibinfo  {journal} {IEEE Transactions on information theory}\ }\textbf
		{\bibinfo {volume} {35}},\ \bibinfo {pages} {197} (\bibinfo {year}
		{1989})}\BibitemShut {NoStop}%
	\bibitem [{\citenamefont {Kuleshov}\ and\ \citenamefont
		{Grudin}(2013)}]{kuleshov2013spectral}%
	\BibitemOpen
	\bibfield  {author} {\bibinfo {author} {\bibfnamefont {E.}~\bibnamefont
			{Kuleshov}}\ and\ \bibinfo {author} {\bibfnamefont {B.}~\bibnamefont
			{Grudin}},\ }\bibfield  {title} {\bibinfo {title} {Spectral density of a
			fractional {B}rownian process},\ }\href@noop {} {\bibfield  {journal}
		{\bibinfo  {journal} {Optoelectronics, Instrumentation and Data Processing}\
		}\textbf {\bibinfo {volume} {49}},\ \bibinfo {pages} {228} (\bibinfo {year}
		{2013})}\BibitemShut {NoStop}%
	\bibitem [{\citenamefont {Sibatov}\ and\ \citenamefont
		{Uchaikin}(2007)}]{Sibatov2007FractionalDifferentialKinetics}%
	\BibitemOpen
	\bibfield  {author} {\bibinfo {author} {\bibfnamefont {R.~T.}\ \bibnamefont
			{Sibatov}}\ and\ \bibinfo {author} {\bibfnamefont {V.~V.}\ \bibnamefont
			{Uchaikin}},\ }\bibfield  {title} {\bibinfo {title} {Fractional differential
			kinetics of charge transport in unordered semiconductors},\ }\href
	{https://doi.org/10.1134/S1063782607030177} {\bibfield  {journal} {\bibinfo
			{journal} {Semiconductors}\ }\textbf {\bibinfo {volume} {41}},\ \bibinfo
		{pages} {335} (\bibinfo {year} {2007})}\BibitemShut {NoStop}%
	\bibitem [{\citenamefont {Alaria}\ \emph {et~al.}(2019)\citenamefont {Alaria},
		\citenamefont {Khan}, \citenamefont {Suthar},\ and\ \citenamefont
		{Kumar}}]{Alaria2019ApplicationFractionalOperators}%
	\BibitemOpen
	\bibfield  {author} {\bibinfo {author} {\bibfnamefont {A.}~\bibnamefont
			{Alaria}}, \bibinfo {author} {\bibfnamefont {A.~M.}\ \bibnamefont {Khan}},
		\bibinfo {author} {\bibfnamefont {D.~L.}\ \bibnamefont {Suthar}},\ and\
		\bibinfo {author} {\bibfnamefont {D.}~\bibnamefont {Kumar}},\ }\bibfield
	{title} {\bibinfo {title} {Application of fractional operators in modelling
			for charge carrier transport in amorphous semiconductor with multiple
			trapping},\ }\bibfield  {journal} {\bibinfo  {journal} {International Journal
			of Applied and Computational Mathematics}\ }\textbf {\bibinfo {volume} {5}},\
	\href {https://doi.org/10.1007/s40819-019-0750-8} {10.1007/s40819-019-0750-8}
	(\bibinfo {year} {2019})\BibitemShut {NoStop}%
	\bibitem [{\citenamefont {Caputo}\ and\ \citenamefont
		{Fabrizio}(2016)}]{caputo2016applications}%
	\BibitemOpen
	\bibfield  {author} {\bibinfo {author} {\bibfnamefont {M.}~\bibnamefont
			{Caputo}}\ and\ \bibinfo {author} {\bibfnamefont {M.}~\bibnamefont
			{Fabrizio}},\ }\bibfield  {title} {\bibinfo {title} {Applications of new time
			and spatial fractional derivatives with exponential kernels},\ }\href@noop {}
	{\bibfield  {journal} {\bibinfo  {journal} {Progress in Fractional
				Differentiation \& Applications}\ }\textbf {\bibinfo {volume} {2}},\ \bibinfo
		{pages} {1} (\bibinfo {year} {2016})}\BibitemShut {NoStop}%
	\bibitem [{\citenamefont {Losada}\ and\ \citenamefont
		{Nieto}(2015)}]{losada2015inverse}%
	\BibitemOpen
	\bibfield  {author} {\bibinfo {author} {\bibfnamefont {J.}~\bibnamefont
			{Losada}}\ and\ \bibinfo {author} {\bibfnamefont {J.~J.}\ \bibnamefont
			{Nieto}},\ }\bibfield  {title} {\bibinfo {title} {Properties of a new
			fractional derivative without singular kernel},\ }\href@noop {} {\bibfield
		{journal} {\bibinfo  {journal} {Progr. Fract. Differ. Appl}\ }\textbf
		{\bibinfo {volume} {1}},\ \bibinfo {pages} {87} (\bibinfo {year}
		{2015})}\BibitemShut {NoStop}%
	\bibitem [{\citenamefont {Estrada}\ and\ \citenamefont
		{Kanwal}(2000)}]{estrada2000singular}%
	\BibitemOpen
	\bibfield  {author} {\bibinfo {author} {\bibfnamefont {R.}~\bibnamefont
			{Estrada}}\ and\ \bibinfo {author} {\bibfnamefont {R.~P.}\ \bibnamefont
			{Kanwal}},\ }\href@noop {} {\emph {\bibinfo {title} {Singular integral
				equations}}}\ (\bibinfo  {publisher} {Springer Science \& Business Media},\
	\bibinfo {year} {2000})\BibitemShut {NoStop}%
\end{thebibliography}
%

\end{document}